# Linguistic Hooks: Investigating The Role of Language Triggers in Phishing Emails Targeting African Refugees and Students

Mythili Menon*
Wichita State University
mythili.menon@wichita.edu

Nisha Vinayaga-Sureshkanth*
University of Texas at San Antonio
vsnisha@ieee.org

Alec Schon
Wichita State University
ahschon@shockers.wichita.edu

Kaitlyn Hemberger†
University of Kansas
k908h677@ku.edu

Murtuza Jadliwala
University of Texas at San Antonio
murtuza.jadliwala@utsa.edu

## Abstract

Phishing and sophisticated email-based social engineering attacks disproportionately affect vulnerable populations, such as refugees and immigrant students. However, these groups remain understudied in cybersecurity research. This gap in understanding, coupled with their exclusion from broader security and privacy policies, increases their susceptibility to phishing and widens the digital security divide between marginalized and non-marginalized populations. To address this gap, we first conducted digital literacy workshops with newly resettled African refugee populations ($n$ = 48) in the US to improve their understanding of how to safeguard sensitive and private information. Following the workshops, we conducted a real-world phishing deception study using carefully designed emails with linguistic cues for three participant groups: a subset of the African US-refugees recruited from the digital literacy workshops ($n$ = 19), African immigrant students in the US ($n$ = 142), and a control group of monolingual US-born students ($n$ = 184). Our findings indicate that while digital literacy training for refugees improves awareness of safe cybersecurity practices, recently resettled African US-refugees still face significant challenges due to low digital literacy skills and limited English proficiency. This often leads them to ignore or fail to recognize phishing emails as phishing. Both African immigrant students and US-born students showed greater caution, though instances of data disclosure remained prevalent across groups. Our findings highlight, irrespective of literacy, the need to be trained to think critically about digital security. We conclude by discussing how the security and privacy community can better include marginalized populations in policy making and offer recommendations for designing equitable, inclusive cybersecurity initiatives.

## 1 Introduction

Phishing is a form of sophisticated social engineering attack in which social engineers exploit human trust in order to trick individuals or corporations into sharing privately identifiable information (PII) through clicking malicious links, downloading infected attachments, and enabling unauthorized access which can be later used for identity theft or financial fraud [1, 4, 6]. Although there is a huge body of literature studying phishing, the focus has been on the Western, Educated, Industrialized, Rich, and Democratic (WEIRD) populations who speak English as their primary language [3]. This focus on a particular kind of demographic excludes nearly 22% of the population in the United States (US) who speak a language other than English as their primary language [9], and often come from non-Western and less industrialized or rich countries. This population is primarily made up of bilingual immigrants and refugees from diverse sociocultural backgrounds, each with different levels of understanding of cybersecurity, privacy, and social engineering attacks [13, 23, 47]. Many come from regions like Africa, where their experiences with technology and digital security may vary widely [14]. As a result, their expectations and awareness of online threats can differ, making them more susceptible to risks such as phishing, identity theft, and other social engineering attacks [24]. These populations already face significant barriers to access (like language or education), making them more susceptible to cyber threats. Therefore, the research gap where studies focusing on non-WEIRD populations are non-existent only widens the digital inequity. It is imperative to understand how this vulnerable population responds to phishing attacks in order to mitigate the harmful effects of PII disclosure and, in turn, develop cybersecurity policies that are more inclusive and equitable.

Phishing combined with inadequate privacy measures for refugees and immigrants can increase their vulnerability to online socially engineered attacks. Targeted spear-phishing attacks, such as those against the Rohingya refugees [49] show that scammers can exploit trust and vulnerability by posing as humanitarian agencies or by sending highly personalized and convincing phishing attempts, targeting personal needs or urgency. Some consequences of phishing attacks include significant financial losses through identity theft or by providing bank account information, security risks while using public devices, such as accessing emails on a computer in the local library, and increased psychological distress in a population that may already have post-traumatic stress disorder (PTSD) and anxiety. Thereby, phishing is a direct threat to privacy and disclosure of PII in an already vulnerable population, such as refugees, who have consistently traded PII for basic needs such as shelter or food.

There is now a growing body of research exploring qualitative perceptions of immigrants and refugees regarding cybersecurity, privacy practices, and their specific needs in these areas. This growing literature seeks to understand how these communities view and understand digital security, the challenges they face, and the gaps in knowledge or resources that may leave them vulnerable to online threats [22, 39, 44]. However, many of these studies have primarily relied on using qualitative methods, such as semi-structured and focus-group interviews. A recent study has shown that there is tremendous potential in the use of quantitative methods, such as measuring privacy inequities and disparities through real-world deception studies [35]. In this paper, we leverage on this observation and ask the question: *How do immigrants and refugees respond to real-world phishing attacks targeting them*? We answer this question by studying how recent US resettled African refugees and African immigrant students studying in the US respond to phishing attacks. These populations are highly vulnerable to phishing attacks due to differences in cybersecurity policies and practices in their home countries, such as the Democratic Republic of the Congo (DRC) [2]. Lack of proper security measures combined with poor digital literacy skills puts this population at high risk and vulnerable to phishing scams (more details in Section 2.3).

---

*Both authors contributed equally to this work.
†Work done while at Wichita State University.

To understand how vulnerable populations such as refugees and immigrant students respond to phishing attacks, we conduct the first comprehensive real-world phishing study on African US-refugees and African immigrant students as outlined in Figure 1. We compare these results with US-born monolingual students to serve as a comparison class for refugees and African immigrant students. First, we conduct digital literacy workshops for African refugee participants who have low general literacy skills and digital literacy skills (see Appendix A for topics/activities covered in the workshop). Next, we conduct a real-world phishing deception study for all study populations to systematically record and study responses to linguistic triggers in curated phishing emails using a psycholinguistic 2 x 2 factorial experimental design described in Section 4. We use language variables that have been shown to correlate with deception texts [31], [32]. More specifically, we attempt to thoroughly address and answer the following specific research questions (RQs) in order to extend our understanding of vulnerable users' interaction with phishing attacks and how they shape security and privacy concerns and policies.

> **[RQ1]** Do digital literacy training through workshops improve cybersecurity and privacy *understanding* among refugee populations?
>
> **[RQ2]** What role do *linguistic cues* play in influencing how people interpret phishing versus legitimate emails?
>
> **[RQ3]** Does *literacy* (general and digital) correlate with a better detection of phishing emails in these target populations?
>
> **[RQ4]** Does a better understanding of cybersecurity and privacy and the ability to detect phishing emails depend on *demographic and cultural* factors?

**Ethics Statement.** Our study was approved by the full committee of the Institutional Review Board (IRB) at Wichita State University. Although participants were consented for Phase I of the study (digital literacy workshops for refugee populations), we requested a waiver of consent for Phase II of the study (phishing study for all study populations), which was granted by the IRB. At the end of the study, we debriefed all participants on the nature of the study and gave them the opportunity to opt out of the study. We shared the results of the study and provided additional materials to help protect participants from similar phishing attacks in educational workshops (Phase III of the study). We took extra care while working with refugee populations. Each member of the study team who interacted directly with the refugee population underwent training from the International Rescue Committee (IRC) of Wichita who was our community partner in recruiting this population. We had three interpreters available during the digital literacy workshops who were fluent in Kiswahili, French, and English and they belonged to DRC. We answered any questions that the refugee population may have had during the period in which we interacted with them. During the workshops, we consistently reminded participants that answering the survey questions was completely voluntary.

**Findings.** Across all groups, we found that linguistic cues, literacy levels, and country of origin significantly impact phishing engagement and susceptibility. We found that although low literacy can be a barrier to securing employment, it can be a blessing in disguise for refugee communities when interacting with phishing emails targeting them. Due to low digital literacy skills and overall low general literacy, refugee participants did not interact with the phishing emails we sent them. With respect to high literacy groups, such as African students and US-born students, they are still susceptible to being phished even though they had overall higher digital and general literacy skills compared to the refugee participants. Our findings reveal *a paradox*: limited English proficiency may shield refugee participants from email-based phishing attempts, yet establishing trust through workshops and interactions may lead them to share PII or personal documents. We also observed that high-literacy participants are not immune to disclosing sensitive data. Based on these insights, we offer recommendations for researchers, resettlement agencies, universities, and policymakers to help non-English speaking communities become less vulnerable to phishing and to promote more inclusive cybersecurity strategies for all.

**Contributions.** In this paper, we make the following contributions:

- We design digital literacy workshops to study and improve the understanding of privacy among recently resettled African US-refugees.
- We conduct a real-world phishing deception study including African US-refugees, African immigrant students, and US-born students, and analyze which carefully manipulated linguistic cues in phishing emails are associated with more data disclosures.
- We develop an understanding of how African populations contrast with US populations (1) in comprehending and responding to phishing emails, (2) whether general and digital literacy play a role in being phished, and (3) whether demographic and sociocultural factors contribute to varying understanding of privacy practices.
- We synthesize lessons for (1) interpreters and NGOs who work with refugees and immigrants, and (2) phishing detection and protection of PII.

- We outline future directions and interventions to help K-12, NGOs, and cybersecurity practitioners create safer and inclusive privacy practices for all.

Code and experimental artifacts are available at https://github.com/menon-lab/phishing.

## 2 Background and Motivation

### 2.1 Refugees

*Refugees* are men, women and children who are unable to return to their country of origin or nationality due to persecution or a well-founded fear of persecution [30]. They face numerous challenges. Many of them experience physical and psychological stress in their home country, as well as upon arrival in the host country, such as US. This can increase their vulnerability to mental health problems, depression, and PTSD. Although fluent in several languages, majority of them do not speak, read, or comprehend the language of the host country. This leads to delays in finding employment, coupled with risks of spear-phishing attacks targeting this population. The resettlement process involves coordination with the US Bureau of Population, Refugees, and Migration. Organizations like the IRC manage the Resettlement Support Center (RSC), which helps refugees prepare their cases for the Department of Homeland Security (DHS) by collecting the necessary personal information for security clearance. Once approved for resettlement, the US government collaborates with the IRC and other national resettlement organizations to help refugees start a new life in the US. Refugees may be resettled in cities with established communities that speak their language or share their culture, or where they have friends or relatives. Factors such as affordability of living and accessibility to healthcare are also considered. The process includes steps such as preparation for travel, arrival in the US, and getting settled in their new home. In FY 2023, US refugees mainly originated from the DRC (30%), Syria (18%), Afghanistan (11%), and Burma (10%) [37].

### 2.2 Digital Literacy

*Digital literacy* is increasingly crucial in today's world, as it empowers individuals to navigate the digital landscape effectively. African US-refugees have poor digital literacy skills [43]. By conducting digital literacy workshops, our goal was to improve our refugee participants' understanding of the digital landscape in the US. Immigrant students from Africa have very similar sociocultural and linguistic backgrounds to African refugees. However, their journey to the US, and their literacy skills are very different. These students chose to come to the US to pursue higher education and they have higher literacy skills in their native language, as well as in English. In addition, they also have high digital literacy skills. Therefore, we recruited participants from two groups, low literacy (refugees from Africa) and high literacy (immigrant students from Africa). Comparing and contrasting these two populations, albeit from Africa, provides insights into the role literacy and education play in improving understanding of social engineering attacks and the phishing landscape.

### 2.3 Motivation for the Study

The main *motivation* for this study is due to the growing evidence that refugee and immigrant populations are at increased risk and highly vulnerable to social engineering attacks for the following reasons: a) *differences in understanding cybersecurity policies and practices in their home countries* − refugees have fled war and persecution in their home countries, such as the DRC or Sudan. Many of these countries do not have strict cybersecurity and privacy policies compared to the US [8], b) *sociocultural differences* − refugees maybe more prone to sharing PII, due to differences in sociocultural practices back in their home country where such information may not be used for malicious intent [7], c) *different expectations and lack of general suspicion* − connected to the sociocultural differences, refugees are not primed for social engineering attacks, i.e. they are not expecting an unsolicited email or text message from a social engineer intending to steal their personal information [39], d) *lack of fluency in language of host country* − refugees, especially those that are resettled in the US, have limited English language proficiency and recently there has been an increase in the complexity of language in malicious email communication [11], e) *lack of awareness and fear stemming from differences in the US legal system* − due to a lack of awareness of social engineering attacks combined with a fear of doing the right thing in a new country, or fear of being deported or not having their legal status approved, refugees may more easily divulge personal information [18].

Refugees who have recently resettled in the US face two primary challenges: a) lack of basic English skills, and b) lack of employment opportunities [27]. In order to address these issues, NGO and refugee resettlement agencies provide classes for their clients to improve their English literacy skills and also hold digital literacy workshops where refugees create job resumes and personal emails that are then used when applying for jobs. While these workshops and education help, refugees are still new to the digital landscape in the US, and they remain highly vulnerable to phishing scams.

### 2.4 Participant Groups

While there are several barriers that recently resettled refugee populations face in the US, some of these barriers are in contrast to the experiences of high literacy African immigrant students. Although these students are also vulnerable to social engineering attacks due to differences in cybersecurity policies and practices in their home countries, they have adequate fluency in the language of the host country, as well as required digital literacy skills required to interact with phishing emails, unlike low literacy refugees. Comparing and contrasting these two distinct groups from Africa based on literacy allows us to get a fine-grained perspective of the roles that general literacy, digital skills and education play in combating social engineering attacks. These two groups differ in digital literacy skills in the following ways: a) immigrant students use email and computers on a daily basis, b) immigrant students are required to take standardized tests, such as ACT, SAT, IELTS, and TOEFL back in their home country before getting admitted into US universities. Whereas refugee populations are a) not comfortable using computers, b) do not check email daily, and c) they are taught how to use smartphones after they are resettled in a new country. In addition to African populations, we included US-born students to study the differences between sociocultural expectations. US-born participants have more awareness of social engineering attacks, such as when a bad actor tries to steal online credentials, due to early access to technology from elementary school days, they speak English natively, and they have high digital literacy skills.

## 3 Related Work
### 3.1 Phishing Scams

In email phishing scams, social engineers steal people's identities and try to manipulate their target by offering or appealing to something that requires urgent attention. Social engineers send deceptive emails to lure their targets to disclose sensitive information. These social engineering attacks [12, 19, 20, 48] often rely on psychological manipulation, but can also leverage linguistic features to achieve their goals. In the context of refugee and immigrant populations, the use of linguistic features in social engineering attacks can be particularly effective, as refugees and immigrants may be unfamiliar with the language of their host country and may lack the resources to verify the accuracy of the information presented to them.

### 3.2 Linguistic Analysis

Several studies have analyzed the language used by scammers in crowdfunding and Nigerian 4-1-9 fraud schemes to understand how they manipulate and persuade their targets. For crowdfunding studies, scammers tend to provide minimal information and fewer typographical errors, with 4.5-4.7 times lower rates than non-phishing emails [38]. Contrary to this, Nigerian fraud emails are riddled with linguistic errors [36]. Scammers also use linguistic strategies in spam emails, including cultural indexicals, interactional roles, and narrative strategies (first, second, and third person narratives), to establish a sense of identity and mutual relationship with their targets [17]. Scammers shift their mode of persuasion in a conversation and tend to favor communication-oriented clauses and personal pronouns, using them to index and position themselves relative to their target [5]. The correlation between psychological triggers and the attack success rate has been well studied in the existing literature [45, 46]. However these works have focused on machine learning approaches to studying phishing. Our study differs in not using any grammatical errors in the phishing emails and investigating the specific role of linguistic triggers in phishing behavior through real-world deception.

### 3.3 Minority Groups

In the UK, a study was conducted on *migrant domestic workers* (MDW) to test threats to their privacy and security from government surveillance, scams and harassment, and employer monitoring. This study was later used to create a digital privacy and security guide to foster security in the lives of MDWs [40]. Another study in the US, studied the computer security and privacy practices and needs of recently resettled refugees in the US. In-depth interviews with case managers and teachers, as well as focus groups with refugees themselves, reveal that refugees rely heavily on technology and their support networks to establish their lives and find jobs, which can make security practices difficult to prioritize. Language barriers, limited technical expertise, and cultural knowledge gaps also present significant challenges to computer security and privacy for refugees. The study concludes with recommendations for the computer security community to better serve the needs of this vulnerable population [39].

Existing literature and work on refugees and immigrants do not specifically focus on investigating user behavior when faced with real-world socially engineered phishing emails on refugee population and immigrant populations from Africa where years of conflict and war have resulted in a diminished understanding of the evolving threat landscape. Since we lack any empirical understanding of how African refugees or African immigrants interact with attackers who employ real-world social engineering techniques such as phishing, this population remains potentially vulnerable to disclosing PII, which in turn could lead to significant personal and financial losses. In addition, we also lack carefully created datasets on refugee and immigrant interactions with phishing emails, preventing research progress in making cybersecurity policies truly inclusive.

## 4 Methodology
### 4.1 Experiment 1: Refugee Digital Literacy Workshops and Phishing Study (Group A)

*4.1.1 Recruitment and Creation of Database.* We initially designed the study to specifically recruit low literacy Kiswahili-speaking refugees from DRC who had been in the US for five years or less. This is because the DRC constitutes the largest refugee resettlement in Wichita, Kansas where the study was conducted. However, this original design neglected the broad-scale diasporic situation of central Africa and areas surrounding DRC. To create a broader educational impact and to accommodate the social needs of the wider central African refugee community at large, we expanded participant selection to additionally sample from nearby countries, including Burundi, Malawi, and Sudan. The inclusion criteria were a) refugees who were recently resettled in the US (under 5 years), and b) they spoke Kiswahili. Low literacy refugee participants were invited to participate in the study through the local chapter of our community partner, the IRC. Participants signed a consent form to participate in the digital literacy workshops, and were compensated $15 per hour. All materials used in the workshop are released publicly. Participant demographics is in Table 1.

*4.1.2 Participant Demographics.* We recruited 48 participants ($M=23, F=25$) from various refugee backgrounds to the digital literacy workshops. The majority (35%) fell into the 25−34 age group, followed by the 35−44 age range (29%). These participants have resettled from DRC, Tanzania, Malawi, South Sudan, and Burundi, and they lived in refugee camps in Rwanda for several years before arriving in the US. In particular, 48% had lived in refugee camps before turning 18, and 17% were born in refugee camps. Regarding their time in the US, 91% had resided in the US for less than one year, 6% for 1 to 2 years and 2% for less than 5 years. Marital data indicated that 50% were married, 89% had children, and most reported having 7−9 children. Of this, 19 participants *($M = 7, F = 12$)* participated in the phishing study.

*4.1.3 Pre-assessments.* Working closely with IRC, we conducted 14 pre-sessions with small groups of African refugees who were part of the English language proficiency class at IRC. Each group ranged from 1 to 7 participants in total. All participants ($n = 48; M =$ 23, $F = 25$) were pre-assessed on three questionnaires. During the pre-session, we worked one-on-one with each participant to collect the language proficiency of each participant, assessed using the *LEAP-Q questionnaire* [25], digital literacy was assessed using the *Digital Skills Questionnaire from vINCI* [41], and we also collected information using an extensive demographic questionnaire (Appendix A.1) that we created in our research center, The Center for Educational Technologies to Assist Refugee Learners at Wichita State University. We always had 2-3 interpreters (from DRC) of Kiswahili and French present during these pre-sessions and the interpreters and the project team translated each question into Kiswahili and recorded their answers. This was done to ensure timely completion, as many of the participants are not fluent in reading or writing, even in their own native languages.

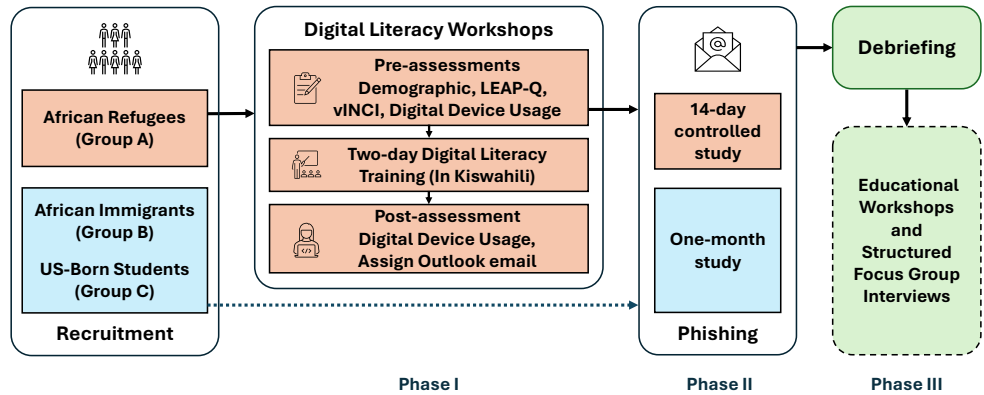

Figure 1: An Overview of the Study Design, Including Recruitment, Assessments, Interventions, and Phishing Across Three Participant Groups. Ongoing Studies are in the Dotted Box.

*4.1.4 Digital Device Usage Questionnaire: Pre-test and Post-test.* Based on the results from the *vINCI* questionnaire, we created a *Digital Device Usage Questionnaire*, which is a six-item Likert-based self-assessment questionnaire of digital device usage and practices. Participants performed this self-evaluation before and after the workshops to assess whether or not their understanding about technologies and their perceived knowledgeability and comfortability had changed. Our pre-assessments showed that African refugee populations were most familiar with operating smartphone devices (as opposed to laptops, computers, or tablets), which became our basis for the modified questionnaire. This enabled us to adapt our digital literacy workshop curriculum to emphasize more on introductory concepts related to operating devices and doing so safely (see Appendix A.4 for the digital device usage questionnaire we created).

*4.1.5 Digital Literacy Workshops.* Pre-assessed refugee participants attended digital literacy workshops held over two days. The goals of these workshops are three-fold: a) to improve English proficiency and literacy of the participants, b) to assess digital skills of the participants, and c) to prime participants for email usage and our real-world deception study. To adhere to the social distancing and COVID-19 protocols previously established at the university and IRC, workshops were conducted over summer 2023 in small focus groups. Each session lasted about two hours. We conducted a total of 14 digital literacy workshops. Although we originally planned for three sessions, we shortened the workshops to two days (sessions) because of several constraints. First, most of our participants relied on public transport, requiring two or three bus transfers to reach IRC where we conducted the workshops. During the summer months, severe heat and infrequent bus routes forced them to miss classes at IRC. Second, most of our participants had small children without a day care option. To maximize the attention span of the participants, we decided that a two-day digital literacy workshop would be most effective.

During the workshops, participants were given hands-on skills and learning opportunities to familiarize themselves with smartphones and privacy practices (see Appendix A for an overview). On day 1, participants completed a modified *digital device usage test*, translated into Kiswahili, before the workshop began. This questionnaire assessed how much they knew about the use of digital devices and the protection of PII and served as a *pre-test survey*. At the end of the second day of the workshop, participants were asked to download the Microsoft (MS) Outlook app on their phone, assigned a university email account, and learned how to compose and send emails. They were encouraged to check their newly assigned email accounts daily to familiarize themselves with the new app and email client. Each participant received a certificate of completion from the university and monetary compensation. One of the main outcomes of the workshop was the creation of a Kiswahili speaking low literacy database that consisted of their phone numbers and their personal Gmail addresses. At the end of the workshop, they were asked to complete the same digital device usage questionnaire, thus completing a *post-test survey*.

*4.1.6 Phishing Study Design for Group A.* Following the workshops, the refugees participated in a *controlled* phishing study. Through their assigned university email account, they initially received emails from our research center (a survey on the digital literacy workshop) and the phishing emails later. However, some could not change their initial passwords (assigned by our University's IT team and common password across all email accounts) and lost access to MS Outlook on their phones. The UI/UX interface of Outlook on mobile devices was difficult for the refugees. This combined with low familiarity with mobile mail clients created delivery and usage challenges for our phishing experiment. We therefore attempted to pivot to Gmail, however, our phishing emails were flagged by Google's spam filters and proved to be unreliable for our controlled study design. Ultimately, given these setbacks, we ended up with a reduced pool of 19 participants (out of 48 participants in the workshops). Phishing emails were sent to this subset (Group A, $n$= 19) through MS Defender [29] using the university email accounts we assigned them. They received one phishing email a day for a span of two weeks, and a total of 12 emails in total. They received an email every single day, except Sunday. This experimental design allowed us to minimize the gap between the training intervention and the real-world phishing study.

**Table 1: Participant Demographics by Age & Country of Origin.**

|  | Group A | Group B | Group C |
|---|---|---|---|
| Workshop participants | $n$ = 48 | - | - |
| Phishing participants | $n$ = 19 | $n$ = 143 | $n$ = 192 |
| Analysis participants | $n$ = 19 | $n$ = 142 | $n$ = 184 |
| 18-24 | 1(5.3%) | 69(48.6%) | 110(59.8%) |
| 25-34 | 7(36.8%) | 48(33.8%) | 31(16.8%) |
| 35-44 | 7(36.8%) | 17(12.0%) | 23(12.5%) |
| 45-54 | 2(10.5%) | 8(5.6%) | 15(8.2%) |
| 55-64 | 0(0.0%) | 0(0.0%) | 4(2.2%) |
| 65-74 | 2(10.5%) | 0(0.0%) | 1(0.5%) |
| United States | 0(0.0%) | 0(0.0%) | 184(100.0%) |
| Kenya | 0(0.0%) | 97(68.3%) | 0(0.0%) |
| Dem. Rep. of the Congo | 16(84.2%) | 21(14.8%) | 0(0.0%) |
| Tanzania | 0(0.0%) | 16(11.3%) | 0(0.0%) |
| Burundi | 1(5.3%) | 2(1.4%) | 0(0.0%) |
| Uganda | 0(0.0%) | 3(2.1%) | 0(0.0%) |
| Rwanda | 0(0.0%) | 2(1.4%) | 0(0.0%) |
| South Sudan | 1(5.3%) | 1(0.7%) | 0(0.0%) |
| Malawi | 1(5.3%) | 0(0.0%) | 0(0.0%) |

## 4.2 Experiment 2: Student Phishing Study (Groups B and C)

*4.2.1 Study Participants.* Group B consisted of African immigrant students ($n$ = 142) studying in the US, while Group C comprised monolingual US-born students ($n$ = 184). Both groups were assumed to be high-literacy based on their admission requirements, as they had to complete standardized tests such as SAT or ACT to be admitted to our university. In addition, international students in Group B had to complete English proficiency tests such as TOEFL, IELTS, or iTEP. Demographic data for both groups was obtained upon request from Wichita State University's Office of Planning and Analysis (OPA), which collects such information from all admitted students. This information is in a publicly available directory for people within the organization and available upon securely signing in, however, students can choose to have their demographic information as confidential and we were not given access to any of that information from OPA.

*4.2.2 Phishing Study Design for Groups B and C.* These groups participated in the real-world deception study over a month. They received phishing emails once a week for four weeks. This design better reflected their routine digital practices, since university students check email frequently and are already primed to encounter phishing attempts in daily life. For each participant, the day and time of delivery were varied each week, and each email corresponded to a different 2 x 2 manipulation condition.

## 4.3 Shared Experimental Features Across Groups

*4.3.1 Linguistic Study Design.* Across all target population groups, participants randomly received one or more emails (outlined in Table 8 in the Appendix) corresponding to four linguistic manipulations determined by two factors − *content framing* and *suspicion level*. These four linguistic manipulations are: (i) **Composite-Mildly Suspicious (*C-MS*)**, ii) **Personal Concern–Mildly Suspicious (*PC-MS*)**, (iii) **Composite–Definitely Suspicious (*C-DS*)**, and (iv) **Personal Concern-Definitely Suspicious (*PC-DS*)**. These scenarios tested how well participants could identify potentially harmful emails, ranging from mild cues of fraud (plausible social or financial appeals) to more obvious red flags (suspicious link, overt demands for sensitive data). The linguistic feature of *composite* inflates the following language variables: analytical thinking, clout, authenticity, and tone. The linguistic feature of *personal concern* inflates the following categories: work and money-related matters. Previous text analysis studies using Linguistic Inquiry and Word Count (LIWC) [33], a software developed for analyzing word use, have shown that fraud and scam texts can be detected using these linguistic features associated with different text analyses [15, 16, 26]. Lower scores of each language variables in LIWC have been shown to correlate with deception texts [31, 32]. We carefully created an email study database with 190 unique stimuli for each factor that we manipulated in our experimental design to elicit a response or reaction in each of our four experimental conditions. We controlled for syntactic length, verb tense, and grammar. There were no typographical or grammatical errors. We used the same length across all emails and we kept the same trusted sender across all experimental conditions and groups- namely the sender was Wichita State University. We computed difficulty level of the emails using the Phish Scale [10, 42] (see Appendix B.1).

*4.3.2 Study Tools.* MS Defender is a comprehensive endpoint protection platform developed by Microsoft that provides robust security solutions, including advanced threat detection and detailed event logging [29]. For our study, we utilized the Attack simulation feature [28] in the Defender portal due to its ability to monitor and record participant interactions with phishing emails in a highly granular manner. The system's logging features allowed us to track every stage of email interaction, providing insights into participant behavior and potential compromises without saving any PII entered by them. Additionally, our University has a phish alert button installed in MS Outlook inbox which lets students, faculty, and staff report suspicious emails to the University's IT team.

*4.3.3 Metrics.* We examined differences in participant responses across all four linguistic manipulation scenarios (*C-MS, C-DS, PC-MS, PC-DS*) and three participant groups. We measured and compared participant interactions in Section 5 using the following key event metrics. **Unread** tracks how often participants did not view or engage with the phishing email at all (marked as *Unread*). **Read** represents emails opened but without further action taken. **Deleted** represents emails opened and then discarded. Both outcomes suggest that the participants exercised some degree of caution or uncertainty, stopping short of disclosure. **Reported** represents participants who flagged the email as phishing using built-in reporting tools, indicating a high level of awareness. **Replied** represents instances where participants responded without surrendering confidential data. **Link Click** represents instances where participants clicked an embedded link but did not finalize a compromise as they refrained from submitting credentials, whereas **Compromised** refers to instances where participants provided private information, such as Date of Birth (DOB) or Social Security Number (SSN).

*4.3.4 Debriefing.* All participants were sent debriefing emails from the lead investigator's official University email, upon completion of the study (see full email in Appendix C). The debriefing email informed participants that they participated in a real-world deception study and they were sent phishing emails to their university accounts. They were given the option of opting-out of the study. A total of 9 participants opted out of the study after the debriefing phase. One participant belonged to Group B (African immigrants), and 8 participants belonged to Group C (US-born students). No participant in Group A chose to opt out. For Group A, emails were sent both to their personal Gmail accounts and their assigned university email accounts. Opted out participant data was deleted from the analysis.

## 4.4 Limitations

Our real-world deception study had several limitations. First, group sizes were imbalanced. While we successfully recruited 48 refugees to our digital literacy workshops, only 19 of them participated in the deception study. By contrast, the immigrant student group and the baseline control group were more comparable in numbers. Second, language is a confounding variable. All phishing emails were sent in English. Most refugee participants had lower literacy skills, which could have made them more cautious when interacting with phishing emails in English. This does not mean they were less vulnerable to phishing in general, since an attack in their native language(s) could pose a higher risk. We note that participants could translate the phishing emails using software installed on their agency-provided phones. During the workshops, most of them primarily used a Kiswahili keyboard on their mobile devices. Third, technical challenges affected the controlled study design in several aspects. Group A was initially given new university-assigned Outlook accounts, but later we switched to sending phishing emails to their existing Gmail accounts provided by the resettlement agency. Google's spam filters blocked many of the phishing attempts, which reduced effectiveness and proved to be unreliable for our controlled study design. In addition, participants may have been less motivated to check or use their newly assigned Outlook accounts, introducing potential bias. Combined with the language barrier, these factors likely lowered overall response rates in refugee groups. Though our study was intended for all age groups, the age distribution of participants was uneven across groups. Groups B and C were composed mainly of traditional university students aged 18–24, while most participants in Group A were in the 25–44 age range. This age imbalance limits how broadly the findings can be generalized. We focused the study on African refugees. We did this to keep the group and their experiences homogeneous. Moreover, the largest refugee group in the US comes from Africa, particularly DRC. However, the refugee diaspora in the US is varied and diverse. Future work can explore other refugee communities and their understanding of cybersecurity and privacy. Similarly, restricting our immigrant student group to only Kiswahili speaking students limits our understanding of user-responses to phishing emails from the broader immigrant student community in the US. Given these limitations, we believe the results of the study offers unique and valuable insights regarding non-WEIRD populations and their interaction with phishing emails.

## 5 Findings

This section reports results for RQ1–RQ4 using quantitative summaries and relevant statistical tests. Group A participated in digital literacy workshops and a two-week controlled phishing study. Groups B and C participated in a one-month long real-world deception study. Group A received intervention and therefore the results are divided separately for Group A. Groups B and C are comparable in number and participated in the same study, so we report their results together.

## 5.1 RQ1: Cybersecurity, Privacy Understanding

*5.1.1 Linguistic Skills.* We found that English proficiency differed significantly within Group A. Specifically, 54% of the participants indicated that they *could speak a little English*, while 44% stated they *could not speak English at all*. Furthermore, 65% said that *they cannot read English well*, and 73% reported having studied only *up to second grade*, with 22% completing elementary school. Consequently, 80% had not learned English in school and had only started learning after age 25 or after having resettled in the US. To gain a more granular view of the linguistic abilities of the participants, we administered the LEAP-Q questionnaire, which assesses literacy levels and daily language usage across different languages (see Appendix A.2). All participants identified Kiswahili as their dominant language and regularly spoke it with family, friends, and on social media. Three participants spoke up to six languages, although the majority spoke two. Almost everyone reported choosing to read in Kiswahili or French, and none felt comfortable answering the LEAP-Q questions regarding English-language proficiency. This finding underscores a strong reliance on Kiswahili or French in day-to-day life and limited adoption of English, regardless of time spent in the US. We found that all participants had received basic Android smartphones upon arrival in the US from IRC, which are equipped with translation software that they use to text their family and friends (in Kiswahili or French).

*5.1.2 Digital Literacy Skills.* We used the *vINCI* questionnaire to measure digital literacy proficiency in areas of communication, content creation, security and problem solving. Most self-reported low confidence in using mobile devices. In fact, apart from two individuals, everyone answered "No" to each question about digital skills (see Appendix A.3), indicating minimal to zero knowledge in multiple areas of digital competence. These findings reveal extremely low levels of digital literacy in this refugee cohort, likely compounded by language barriers and limited prior exposure to technology.

*5.1.3 Pre- and Post-test Results.* We assessed digital device usage (pre-test) and compared the means to the same questionnaire as a post-test. Each participant answered each question in the questionnaire using a Likert scale of 1-5 with 1 being strongly disagree and 5 being strongly agree. Statistical analysis in Table 6 revealed a significant difference between the pre-test ($M = 2.8$, $SD = 0.66$) and post-test ($M = 3.6$, $SD = 0.75$) measures of digital device usage ($t(4) = -4.39$, $p = 0.01$). As summarized in Table 7 found in Appendix, participants demonstrated a stronger grasp of concepts such as password hygiene, safe link-clicking practices, and general caution when sharing personal documents or financial details. This difference in overall means across all 5 items of the digital device usage questionnaire suggests that focused workshops **in the language of low literacy groups**, can help bridge critical gaps in digital literacy for refugee populations.

**Table 2: Click and Compromised Counts and Rates Based on Total Emails Delivered for All Groups, Manipulations and NIST Phish Scale Detection Difficulty Ratings (V = Very Difficult; M = Moderately Difficult).**

| | | Click Rate | | | | Compromised Rate | | | |
|---|---|---|---|---|---|---|---|---|---|
| | | C-DS | C-MS | PC-DS | PC-MS | C-DS | C-MS | PC-DS | PC-MS |
| A | V | 2/57 (3.51%) | 0/19 (0.00%) | 1/57 (1.75%) | 0/19 (0.00%) | 0/57 (0.00%) | 0/19 (0.00%) | 0/57 (0.00%) | 0/19 (0.00%) |
| | M | – | 1/38 (2.63%) | – | 0/38 (0.00%) | – | 0/38 (0.00%) | – | 0/38 (0.00%) |
| B | V | 10/142 (7.04%) | 11/142 (7.75%) | 10/142 (7.04%) | 5/142 (3.52%) | 7/142 (4.93%) | 3/142 (2.11%) | 3/142 (2.11%) | 2/142 (1.41%) |
| C | V | 30/184 (16.30%) | 11/184 (5.98%) | 20/184 (10.87%) | 7/184 (3.80%) | 17/184 (9.24%) | 1/184 (0.54%) | 7/184 (3.80%) | 0/184 (0.00%) |

*5.1.4 Observed Privacy Practices.* During the digital literacy workshops, we found that the participants continued to show a poor understanding of *security and privacy*. For example, many willingly provided the study team with their **refugee travel documents** when completing the questionnaires, although we explicitly stated that private records should remain confidential. This behavior illustrates deep-rooted vulnerabilities and the lack of familiarity with privacy norms within this community, underscoring the need for ongoing, culturally tailored interventions that go beyond a single workshop series.

> **RQ1 (Refugee competencies):** We found that digital literacy workshops improved post-test means in the digital device usage questionnaire and resulted in improved understanding of safer cybersecurity practices for African refugee populations who have recently resettled in the US.

## 5.2 RQ2: Role of Linguistic Triggers

*5.2.1 Scenarios and Engagement.* In the controlled study, Group A read 32.0% of emails received (73/228) emails (See Table 3). The read emails correspond to content that focuses on *urgent relief* or *official updates*. Despite opening these emails, participants rarely clicked the embedded links, suggesting they scrutinized the content before acting or did not understand the content of the email. In contrast, for the second study, Table 4 (Group B) showed a more polarized outcome given 74.5% (423/568) of the delivered emails were unread. However, once participants found a message credible (25.5% were opened), they often took immediate next steps where embedded links were click in 6.3% of cases (36/568). Furthermore, from Table 5 (Group C), we observed a relatively higher rate of engagement, with 55.1% of emails opened (406/736), reflecting the participants' desire to at least check emails that felt relevant, such as *bank alerts* or *potential job opportunities* in particular. Across Groups B and C, approximately 3.1% of all emails received (40/1304) by participants in both groups were compromised.

**Table 3: Group A: African Refugees Email Summary.**

| Response | PC-MS | PC-DS | C-MS | C-DS | Total |
|---|---|---|---|---|---|
| Unread | 37(64.9%) | 39(68.4%) | 40(70.2%) | 39(68.4%) | 155(68.0%) |
| Read | 20(35.1%) | 18(31.6%) | 17(29.8%) | 18(31.6%) | 73(32.0%) |
| Deleted | 3(5.3%) | 3(5.3%) | 4(7.0%) | 3(5.3%) | 13(5.7%) |
| Reported | 0(0.0%) | 0(0.0%) | 0(0.0%) | 0(0.0%) | 0(0.0%) |
| Forwarded | 0(0.0%) | 0(0.0%) | 0(0.0%) | 0(0.0%) | 0(0.0%) |
| Link Clicked | 0(0.0%) | 1(1.8%) | 1(1.8%) | 2(3.5%) | 4(1.8%) |
| Compromised | 0(0.0%) | 0(0.0%) | 0(0.0%) | 0(0.0%) | 0(0.0%) |
| Replied | 3(5.3%) | 2(3.5%) | 3(5.3%) | 2(3.5%) | 10(4.4%) |
| **Delivered** | 57(100.0%) | 57(100.0%) | 57(100.0%) | 57(100.0%) | 228(100.0%) |

*Note: C* and *PC* denote Composite and Personal Concern linguistic cues, whereas *MS* and *DS* denote Mildly and Definitely Suspicious legitimacy cues. Each participant received three emails across each manipulation over the entire study period.

We ran a multinomial logistics regression across all groups with significant effects of manipulation condition across multiple outcomes (See Table 15 in Appendix). For email click, *PC-DS* predictor was not significant ($OR = 0.72$, $p = 0.207$), with all other manipulations significant (*C-DS* ($OR = 0.14, p < 0.001$), *C-MS* ($OR = 0.52, p = 0.018$), and *PC-MS* ($OR = 0.27, p < 0.001$)). For replied, only *C-DS* was significant ($OR = 0.01, p < 0.001$). For emails reported, *C-DS* again was the only significant predictor ($OR \approx 0$, $p < 0.001$), and for read email, no significant effects were found. Overall, we found that urgency and official-sounding emails consistently influenced whether participants opened and read phishing messages in both the controlled two-week study with Group A and the one-month study with Groups B and C. Specifically, subject lines such as "*ACT NOW*", "*FINAL NOTICE*", or "*Your Account Is Expiring*" prompted higher open rates, as participants often worried about missing critical updates or incurring penalties, as seen in Table 8 in the Appendix.

**Table 4: Group B: African Immigrant Students Email Summary.**

| Response | PC-MS | PC-DS | C-MS | C-DS | Total |
|---|---|---|---|---|---|
| Unread | 107(75.4%) | 104(73.2%) | 108(76.1%) | 104(73.2%) | 423 |
| Read | 35(24.6%) | 38(26.8%) | 34(23.9%) | 38(26.8%) | 145 |
| Deleted | 15(10.6%) | 11(7.7%) | 12(8.5%) | 9(6.3%) | 47 |
| Reported | 0(0.0%) | 0(0.0%) | 1(0.7%) | 0(0.0%) | 1 |
| Forwarded | 0(0.0%) | 0(0.0%) | 0(0.0%) | 0(0.0%) | 0 |
| Link Clicked | 5(3.5%) | 10(7.0%) | 11(7.7%) | 10(7.0%) | 36 |
| Compromised | 2(1.4%) | 3(2.1%) | 3(2.1%) | 7(4.9%) | 15 |
| Replied | 1(0.7%) | 0(0.0%) | 2(1.4%) | 0(0.0%) | 3 |
| **Delivered** | 142(100.0%) | 142(100.0%) | 142(100.0%) | 142(100.0%) | 568 |

*Note: C* and *PC* denote Composite and Personal Concern linguistic cues, whereas *MS* and *DS* to Mildly and Definitely Suspicious legitimacy cues. Each participant received one email across each manipulation over the entire study period.

We also found that scenario plausibility, the degree to which an email mirrored real-life concerns, was essential to drive engagement.

**Table 5: Group C: US-born Students Email Summary.**

| Response | PC-MS | PC-DS | C-MS | C-DS | Total |
|---|---|---|---|---|---|
| Unread | 86(46.7%) | 85(46.2%) | 90(48.9%) | 69(37.5%) | 330 |
| Read | 98(53.3%) | 99(53.8%) | 94(51.1%) | 115(62.5%) | 406 |
| Deleted | 51(27.7%) | 39(21.2%) | 49(26.6%) | 30(16.3%) | 169 |
| Reported | 4(2.2%) | 3(1.6%) | 2(1.1%) | 1(0.5%) | 10 |
| Forwarded | 0(0.0%) | 1(0.5%) | 0(0.0%) | 3(1.6%) | 4 |
| Link Clicked | 7(3.8%) | 20(10.9%) | 11(6.0%) | 30(16.3%) | 68 |
| Compromised | 0(0.0%) | 7(3.8%) | 1(0.5%) | 17(9.2%) | 25 |
| Replied | 3(1.6%) | 5(2.7%) | 4(2.2%) | 4(2.2%) | 16 |
| **Delivered** | 184(100.0%) | 184(100.0%) | 184(100.0%) | 184(100.0%) | 736 |

*Note: C* and *PC* denote Composite and Personal Concern linguistic cues, whereas *MS* and *DS* to Mildly and Definitely Suspicious legitimacy cues. Each participant received one email across each manipulation over the entire study period.

For example, in Table 8, *PC-DS* scenarios (Y01-Y12) referenced employment and benefits, often requesting personal or financial details. The *C-MS* scenarios (G01-G12) focused on housing, healthcare or scholarships, appealing to everyday financial or academic needs, while the *C-DS* (P01-P12) and *PC-MS* (B01-B12) scenarios featured urgent billing or job offers, prompting link clicks from participants. This could be because Group B and C participants are more embedded in campus communication systems, making institutional-looking phishing appear normal and authentic, thereby prompting higher engagement rates. *C-DS* manipulation had the most number of compromises in Groups B and C (7/142 (4.9%) and 17/184 (9.2%), respectively), followed by *PC-DS*. As shown in Figure 2, *MS* susceptibility cues received the least number of compromises with no compromises for *PC-MS* phishing emails for Group C. We conducted a Mann-Whitney U test to analyze the impact of the different stimuli in each of the linguistic manipulations. As shown in Table 11, we found that both linguistic cue and legitimacy cues were significant predictors on whether PII was disclosed. For linguistic cues, *C* scenarios had more disclosures than *PC* scenarios ($U = 217,768$, $p = 0.010$). For legitimacy cues, *DS* scenarios had more disclosures than *MS* scenarios ($U = 221,680$, $p < 0.001$).

*5.2.2 Scenarios and Susceptibility.* We examined whether participants who engaged with phishing emails provided PII disclosures, and in turn were compromised. In most cases, opening a phishing email did not equate to disclosure; rather, linguistic cues and triggers often determined whether a participant disclosed PII. In the first study, Group A, based on Table 3, had four instances of *Link clicked* but zero *Compromises*. We found that while language demonstrating urgency in the email subject (such as *CLAIM YOUR BENEFITS*) enticed them to read, they they may have non-engaged with the contents due to language and comprehension barriers. Low engagement and zero compromises reflected in non-disclosure of PII. This approach was reflected in the scenario-level data: *C-MS* (housing or scholarship offers) and *PC-DS* (job or financial benefits) scenarios elicited *Read* responses but not deeper engagement or PII disclosure.

We also observed that Group B participants opened 25.5% of emails (145/568 in Table 4), and in 6.3% of cases (36/568) proceeded to click a link if the message appeared legitimate to them. Of these clicks, 41.7% (15/36) resulted in compromised credentials, indicating high chances of susceptibility once initial hesitation was overcome. Job-oriented emails or localized references, such as *"University requires your birthdate"* were especially convincing. For compromised, all predictors were associated with significantly reduced odds (*C-DS* ($OR = 0.08$, 95% CI [0.05, 0.12], $p < 0.001$), *C-MS* ($OR = 0.16$, 95% CI [0.05, 0.46], $p = 0.001$), *PC-DS* ($OR = 0.40$, 95% CI [0.19, 0.85], $p = 0.017$), and *PC-MS* ($OR = 0.08$, 95% CI [0.02, 0.33], $p = 0.001$).

Group C participants recorded 9.2% link clicks (68/736) and 3.4% compromises (25/736 in Table 5) with respect to the total emails they received with a 36.8% click-to-compromise ratio (25/68 from Table 2). In contrast, Group B participants had fewer clicks overall (6.3%), but once they clicked, they were slightly more likely to compromise, with a 41.7% click-to-compromise ratio (15/36). This finding indicates a higher likelihood of compromise after clicking email links across both groups (B and C). We found that standard corporate or financial motifs (e.g., *"Credit Card Bill"* and *"Payment Due"*) were particularly persuasive. Across all groups, *C-DS* manipulations were compromised the most, as seen in Figure 2. Participants provided PII disclosures to emails that requested payment or tuition related issues. A two-way ANOVA was run to investigate the main effects of linguistic manipulation cues in Groups B and C. As seen in Table 17, the analysis revealed a significant main effect of manipulation type ($F(3) = [10.42]$, $p < 0.001$). There was no main effect of group membership and no interaction effects. We ran a Tukey's HSD post hoc test which revealed that the *C-DS* manipulations resulted in higher credentials being supplied (Table 14).

> **RQ2 (Linguistic Triggers):** Across all participant groups, we found that emails that emphasize urgency, or connected to the university related processes generated the most engagement. Phrases like "Enter your SSN now" or "Verify your bank details immediately" proved potent, especially when combined with legitimate-seeming language (e.g., university staff references, local affiliations). DS emails received the most compromises, suggesting participants were not hesitant to click suspicious links.

### 5.3 RQ3: Literacy Levels

*5.3.1 Literacy and Engagement.* We hypothesized that lower general literacy skills of Group A could suggest higher vulnerability, however, as seen in Table 3, we found that the MS Defender tool recorded zero successful phishing disclosures. Participant engagement was lower probably due to language and comprehension barriers. For example, participants often asked clarifying questions before any further steps, exemplified by the ten *Replied* entries, which suggests that while they lacked advanced reading or digital fluency, they compensated with skepticism or verification-seeking behaviors. A Spearman's correlation was conducted to evaluate the relationship between literacy and emails read. There was a small significant effect between literacy level and emails read, $\rho_s = 0.157$ with $p = 0.0176$.

In contrast, Group B's advanced general literacy skills in English and Kiswahili led to fewer overall reads equivalent to nearly one-fourth of delivered emails (145/568, 25.5%) but a much higher rate of link clicks once an email passed their initial filter in 24.8% of read emails by the group (36/145). Specifically, 6.3% of delivered emails (36/568) resulted in clicks, and 74.5% (423/568) were left unread, reflecting stronger screening or skepticism. Meanwhile, Group C showed substantially more engagement, opening 55.1% of delivered emails (406/736). They also displayed protective behaviors, with

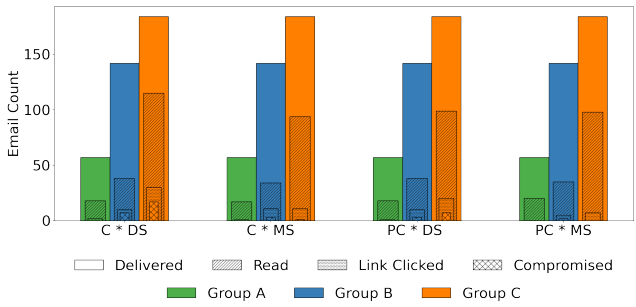

**Figure 2: Phishing Simulation Results by Manipulation Scenario and Group Across Both Studies.**

23.0% of emails deleted (169/736) and 1.4% reported (10/736). At the same time, they engaged in risky behaviors such as clicking phishing link clicks with 9.2% of emails (68/736). A binomial logistic regression was conducted to examine whether reading proficiency scores predicted the likelihood of supplying credentials, clicking on definitely suspicious email links, replying to the phishing emails, deleting emails, reading emails, and reporting emails (See Table 16 in Appendix). The overall model was statistically significant for emails replied ($OR$ = 4.19, 95% CI [1.21, 14.43], $p$ = 0.023), indicating that reading proficiency level reliably distinguished between those who did and did not reply to the phishing emails. Participants with a reading score of 5 were significantly more likely to reply to an email. The model was not significant for supplying credentials, clicking on definitely suspicious email links, deleting emails, reading emails, and reporting emails suggesting that general literacy levels was not a predictor of these outcomes.

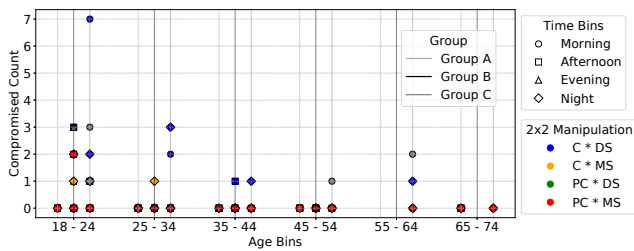

**Figure 3: Breakdown of Compromised Participants by Age, Manipulation, Time and Group in Both Studies.**

*5.3.2 Literacy and Susceptibility.* When participants went beyond clicking and actually provided PII or other information, we noticed that higher literacy did not guarantee greater security awareness. Across all studies, while Group A (Table 3) recorded zero compromises, Group B (Table 4) had the highest click-to-compromise ratio of 41.7% with 15 compromises resulting from 36 link clicks compared to Group C which recorded 25 compromises out of 68 clicks (36.8%). However, the binomial logistic regression model was not statistically significant for credential supplied/compromised ($OR$ = 1.30, 95% CI [0.68, 2.48], $p$ = 0.434) and only marginally significant for *DS* click links ($OR$ = 1.50, 95% CI [0.99, 2.29]), $p$ = 0.057). Participants with stronger reading abilities were more likely to click links on emails that had a job or scholarship-themed content from a trusted sender, such as Wichita State University. The relationship between language proficiency and context highlights a major risk: those confident in detecting scams may still be tricked by emails that imitate known cultural or linguistic patterns, indicating that perceived authenticity often overrides reading ability. Therefore, literacy levels, measured by reading skills or formal education, are not the only indicators of phishing susceptibility. Factors such as *local language, trust in specific institutions, and previous experiences with education or scams* can be just as crucial.

*5.3.3 Literacy and Timing.* Week-by-week email interactions in Table 18 reveal that literacy and speaking proficiency levels were not a consistent safeguard against phishing attempts over time. Across both Groups B and C, over time clicks stayed roughly the same (See Figure 4). This trend suggests that despite higher literacy levels and better English-speaking proficiency, phishing emails continued to provoke risky behavior in participants when the content or context resonated with their daily concerns. Furthermore, we also observe steady link click rates throughout the weeks. Trends of email interaction across the weeks suggest that factors such as workload, time of day, or stress levels may be at play. This suggests that phishing attacks are often more sophisticated than just being a "recognition" problem, it is also about decision making under pressure or confusion.

> **RQ3 (Literacy Correlation):** We found that higher general literacy skills did not always translate into lower susceptibility to phishing. The higher literacy skills of Group B (than those of Group A) did not prevent them from being compromised by scenario-specific details, while the cautious stance of Group A helped them avoid any compromises entirely. This outcome highlights the importance of context and suspicion rather than reading or digital literacy skills alone.

### 5.4 RQ4: Demographics

*5.4.1 Age and Susceptibility.* We examined age-related patterns in phishing susceptibility and, as shown in Figure 3, observed that participants aged 18–24 consistently emerged as the most engaged age group across all participant groups. These young adults also received the highest volume of phishing emails and also demonstrated relatively high engagement rates which was statistically significant ($\chi^2$ = 14.1460, df = 5, $p$ = 0.01470). We further analyzed compromised rates by age group and observed that Group B participants in the same age group disclosed information across different phishing content scenarios resulting in 6 compromises in the *C-DS* scenarios (8.7% of 69 *C-DS* emails delivered to the age group) and 3 compromises in the *C-MS* scenarios (2.9% of 69 *C-MS* emails delivered to the age group). However, the difference is not statistically significant ($\chi^2$ = 6.8152, df = 3, $p$ = 0.078028). Comparatively, younger adults in Group C also disclosed information, though at slightly lower levels than their Group B counterparts, where 9.1% of delivered *C-DS* emails (10/110) and 3.6% of delivered *PC-DS* emails (4/110) to that group resulted in a compromise. However, the difference is not statistically significant ($\chi^2$ = 10.4772, df = 5, $p$ = 0.062790).

In contrast, participants in the 35 to 44 age bracket exhibited minimal disclosures with respect to the emails they received overall (1/68, 1.5% in Group B; 1/92, 1.1% in Group C), while those aged 45 to 54 recorded only one compromise in Group C (1/60, 1.7%). In contrast, participants aged 55 to 64 years recorded 3 compromises in Group C (3/16, 18.8%), highlighting that older adults may be more likely to disclose PII on phishing attempts when scenarios are relevant or personally meaningful. No disclosures were recorded for participants in the 65-74 age range across all groups. Moreover, Figure 3 shows that 18-24 year old immigrant students from Africa had more compromises during the afternoon (12PM-5PM), with minimal to no compromises in the evenings (5PM-9PM) or nights (post 9PM). This pattern highlights the importance of considering participants' daily schedules and routines in assessing phishing susceptibility. However, the difference is not statistically significant ($\chi^2 = 1.6518$, df = 3, $p = 0.647692$). One plausible explanation based solely on the available results is that younger adults often manage multiple digital accounts (academic, professional, or social), which could increase their exposure to phishing. At the same time, they might not yet have the depth of real-world experience needed to thoroughly detect fraudulent cues. Specifically, they often navigate multiple digital services (e.g., job boards and financial apps), matching many of the urgent or *DS* scenarios (*PC-DS* or *C-DS*) in Table 8 in the Appendix. In summary, while the highest overall susceptibilities were observed in the youngest adult cohort (age range 18-24), participants in all age brackets can be targeted with sufficiently tailored phishing messages, underscoring the importance of linguistic cues over purely demographic characteristics.

*5.4.2 Country and Engagement.* We found that country of origin differences were overshadowed by the consistency of well-tailored phishing messages. As seen in Table 3, in Group A, no credentials were compromised irrespective of participants' country of origin. However, Malawi participants engaged more with phishing emails, with 16.7% of their delivered emails resulting in clicks in the *C-DS* condition (2/12) and 8.3% in the *PC-DS* condition (1/12) which was statistically significant ($\chi^2 = 9.4212$, df = 3, $p = 0.024185$). This aligns with previous findings on the relationship between literacy levels, scenario, and cautious behavior. In Group B, although citizens from some countries showed small or large compromise rates in Table 19, these variations appeared to depend on how closely phishing scenarios matched local norms or concerns. For example, Kenyan citizens, with 372 emails, recorded two compromises (0.54%), while Tanzanian citizens (68 emails) had six compromises (8.82%). We also found that Ugandan citizens (16.7% compromised; 2/12 emails), Rwandan citizens (12.5% compromised; 1/8 emails), Burundian citizens (14.3% compromised; 1/7 emails) had higher *compromised* percentages compared to the total emails sent. This was statistically significant ($\chi^2 = 16.4790$, df = 7, $p = 0.021083$). Such contrasts suggest that simple email volume does not dictate engagement outcomes, rather, a combination of cultural familiarity with scam tactics and scenario specificity is what drives individuals to read, click, or ultimately disclose information. However, once we account for the linguistic group of each participant and the relevance of the scenario, the country of origin alone did not emerge as a clear predictor of the phishing results.

> **RQ4 (Demographics):** We found that both age and country of origin served as a predictor of phishing susceptibility. Younger adults in their early 20s engaged more due to heightened job and school demands.

## 6 Qualitative Responses

We invited all participants in our study to voluntarily participate in a qualitative follow-up interview. We interviewed 10 participants (1 from Group A, 3 from Group B, 6 from Group C). We summarize some of the relevant responses in Table 9 in the Appendix. Most participants had been scammed before. Many of the participants ignore emails with promotions or package delivery click links because they said they do not shop online or get packages delivered to their houses. Among this group, we had two participants who disclosed personal information. They responded to definitely suspicious phishing emails with clickable links. Both emails had the university name in the subject line and they offered a job or a payment plan. The landing page after they clicked, spoofed our university's landing page, thereby making it indistinguishable and authentic to the participants. All participants said more awareness and resources are needed to help protect people from scams. Participant B1, from Group B, mentioned mandatory cybersecurity training for all students should be a requirementt in universities.

## 7 Discussion

We discuss key observations, challenges encountered in the study and share lessons learned. We also offer technical and research recommendations aimed at better supporting non-English speaking populations in understanding and navigating cybersecurity and privacy issues.

### 7.1 Privacy Equity

Our findings highlight the important role that legitimacy and linguistic cues play in making decisions about privacy and cybersecurity. In our controlled, real-world deception study with African US-refugee populations, we found that participants with limited English skills or low digital literacy are more likely to ignore or fail to respond to phishing emails. They may not have fully grasped the nuances of the phishing attempt or may have struggled to comprehend an unfamiliar language. In contrast, when studying African immigrant populations and US-born students, we discovered that being literate did not necessarily equate to a better understanding of cybersecurity or privacy risks. Interestingly, data disclosures were often linked to how authentic the message appeared to the participant. This sheds light on how cognitive biases lead participants to depend heavily on heuristics, like the perceived normality or official appearance of a message when judging its authenticity. In addition, there is a gap in understanding what digital literacy entails beyond just reading comprehension. Although high literacy populations can read well, they have not been trained to think *critically* about digital security. Being trained to think critically about security involves questioning the legitimacy of communications, recognizing subtle linguistic cues in emails, and understanding how social engineering works. This training should be a fundamental part of any security curriculum.

Our findings also highlight that awareness and education need to go beyond just identifying threats: at risk and vulnerable users need to cultivate a mindset of caution. They underscore the need for privacy-enhancing technologies that are deliberately inclusive in their design. Systems that assume fluency, technical competence, or regular digital access risk reinforcing existing inequities by failing to protect those most vulnerable to exploitation. Our study highlights several avenues for improvement, including multilingual and visually guided privacy warnings, simplified authentication processes that minimize reliance on complex text, and interfaces that deliver clear and actionable feedback. By framing phishing vulnerability as a challenge of privacy equity, this work identifies opportunities for future technologies to provide meaningful protection across diverse user groups rather than concentrating safeguards among those with the highest levels of digital literacy. Future research should extend this lens to other domains of online harm, such as misinformation, surveillance, and financial fraud, in order to ensure that inclusive privacy protections are informed by a broader understanding of risk.

### 7.2 Challenges to the Study

One significant challenge we encountered during the study was Gmail's robust spam filtering and anti-phishing measures. Given that Group A had significant difficulties accessing the newly assigned Wichita State University MS Outlook email address through their phone app, we pivoted to sending them phishing emails to their personal IRC-provided Gmail addresses. Although our initial plan involved using custom email servers to send these study-related phishing emails directly to the personal Gmail accounts of participants in Group A, these emails were quickly identified and blocked by Gmail's phishing detection systems. As a result, participants never received (or even saw) the majority of our phishing emails, rendering the study ineffective. Attempts to adjust server settings or sender details did not succeed, as Gmail continued to flag the messages as suspicious, likely due to key terms used in the email body such as *request for SSN*, etc. This change and challenge in sending emails to participants with low digital literacy skills impacted the results we had hoped to receive for Group A.

### 7.3 Lessons Learned

Our findings highlight the need for access to *bilingual* resources on cybersecurity and privacy practices for non-English speaking populations. These resources should be basic, visually appealing, without any jargon. The results of our digital literacy workshops for low literacy refugee populations sheds light on additional training, feedback, and technology exposure that can help protect this vulnerable population against targeted technology-based scams in an ever-evolving risk landscape. We identified a need for more tailored, localized, and language-specific initiatives.

> Interpreters are essential to the success of digital literacy workshops.

Many NGOs working with refugee resettlement, such as IRC, are short-staffed and rely on volunteers. Staffing digital literacy workshops can be difficult because refugees in these classes speak a variety of languages. Thus, future researchers working with refugee populations should aim to group refugees into language groups with multiple interpreters. Having multiple interpreters in the digital literacy workshops that we conducted enabled feedback loops where clarification was sought and received in a language most comfortable for the refugees to understand. This resulted in better retention of workshop material as seen by significant change in post-test means.

> Consistent with our hypotheses, recently resettled refugees (under 5 years of moving to US) with low literacy have a rudimentary understanding of cybersecurity and privacy.

For example, during the pre-sessions, when our study team asked for the ages of the children of the participants, some of the refugee participants handed out their *refugee travel document*, which is the document they use to enter the US in lieu of passports. This document is a sheet of paper with photographs and PII (such as DOB, travel document number) of each family member accompanying the head of the household.

> Established trust creates a safe space for the refugee participants to share the document without fear of the PII being misused

Many refugee participants do not remember their DOB or passwords, and they keep them written down on a piece of paper in their pocket, wallet, or bag. DOB is culturally not as relevant in rural Africa where they grew up, or in refugee camps. Some of them, in fact, have made-up DOB for resettlement purposes. They have frequently exchanged PII such as DOB with many officials, both in their home country or refugee camps and in the US, for refuge and shelter. This increases their risk of being harmed by phishing attacks trust is established with the social engineer.

> Low literacy skills correlates with low digital literacy skills.

Typing on a phone keyboard is very hard and is correlated with age, education, and gender. Moreover, typing in English is even harder and a very slow process which becomes a major barrier for refugees who are trying to find a job in the US, as most jobs require online applications. When we asked participants to type their new email address in the MS Outlook app on their phones, many of them were unable to do so or had to resort to Kiswahili or French keyboards. The UI/UX interface of MS Outlook made it very difficult for many of our participants to successfully interact, receive, and send emails on MS Outlook.

> Lack of English language proficiency may be a shield for refugee communities protecting them from phishing attacks.

The refugee community is more likely to ignore an email when it is in English (or a language unfamiliar to them), which reduces the chances of them opening, responding to, or engaging with the malicious content. This behavior was evident in the real-world controlled phishing study.

### 7.4 Recommendations

Proficiency in the language of the host country is directly related to an improved understanding of cybersecurity and privacy. We call for better cybersecurity educational modules focused on non-English speaking populations. However, these modules cannot be a *one size fits all* model as population needs differ depending on their native language and other languages these communities speak.

> ⚠ Developers should focus on creating games that could teach vulnerable populations about the costs of social engineering attacks.

These games should be bilingual, at a minimum, to aid in better retention of the material. Creating visual resources with minimal text can be an effective way to mitigate the language barrier. All texts should be translated into appropriate languages and made available to communities. Games have been shown to be a safe space for refugee communities. Some phishing training tools such as Proofpoint security awareness training [34] and KnowBe4 [21] exist, but they need to be adapted to non-WEIRD populations.

> ⚠ Cybersecurity and privacy education should start as early as K-12.

Good cybersecurity practice begins with better awareness of the threat landscape. Interactive sessions in middle and high school led by cybersecurity researchers and practitioners should be implemented. Many refugee participants with whom we worked were helped by their school-going children in operating digital devices, as they learn English in school. Many refugee children in public schools will therefore benefit from being aware of existing scams and social engineering attacks. In turn, they can warn their parents and help their family establish safe cybersecurity and privacy practices. We conducted a one-day workshop on cybersecurity with refugee children who were in high-school as part of our NGO community partner's summer camp. The refugee children began with little to no understanding of cybersecurity. During the workshop, we gave the students a hands-on task called *sorting the messages*. In this game, the students were asked to figure out whether an email was a phishing email or not. We also advised the students to inform their parents and family members about digital threats, as the students mentioned that their families were not fully aware of these risks. It is also important to educate young children about the importance of individuals' right to privacy and the kind of privacy threats that exists in systems that students interact with in their daily lives.

> ⚠ Universities should design more customized and controlled mock phishing attacks to better simulate real-world threats

Current phishing email models used for educational purposes in universities fail to account for the diverse linguistic and socio-cultural backgrounds of students. We urge universities to develop phishing training simulations that consider user susceptibility based on these factors. For instance, First-Generation students may have different expectations or understandings of cybersecurity and privacy. By creating targeted phishing simulations tailored to various demographics, universities can improve awareness and reduce students' vulnerability to phishing attacks.

> ⚠ More efforts should be directed toward understanding and addressing security and privacy needs and practices of non-WEIRD (Western, Educated, Industrialized, Rich, and Democratic) and vulnerable populations

We urge researchers to conduct empirical studies on non-WEIRD populations. Refugees are underrepresented minorities in STEM and their successful integration into US society is important for their overall success in society. Improved understanding of the nature of social engineering threats facing this population will inform policymaking, lead to enhanced social and economic security, and close the inequity gap between non-vulnerable populations and refugees. Studying differences in linguistic triggers cross-linguistically is a fruitful area of research. Tailoring cybersecurity education and awareness programs to these populations will help reduce their vulnerability and ensure that cybersecurity practices are inclusive and accessible to all.

## 8 Future Work

We have ongoing educational workshops and qualitative interviews (Phase III) to understand how participants responded to emails in our study. A summary of the results have been provided in Section 6. During the workshop, participants receive study materials and are educated on how to protect their sensitive information from social engineering attacks. Participants also receive a phishing checklist infographic in their native language, and are asked to distribute this with family and friends. This infographic will be freely available to humanitarian organizations and resettlement agencies. Future studies can build on the limitations and challenges we faced by conducting the study in a language that the refugee population is proficient in, such as Kiswahili or French. Studies should also provide additional practice with e-mail clients, especially if assigning new email addresses to participants. Future studies should also consider working with older populations in order to avoid the younger population bias we had in the current study.

## 9 Conclusion

Although phishing is well studied, the impact of real-world phishing targeting non-English speaking populations such as refugees and immigrants is understudied. To address this gap, we first conducted digital literacy workshops with African US-refugees followed by a real-world deception study on African US-refugees, African immigrant students, and US-born monolingual English-speaking students. Our study reveals that lower general literacy and digital literacy skills among African US-refugees may inadvertently protect against email phishing, while trust built through interpersonal interactions can increase vulnerability, suggesting that more intervention is required for this population. In contrast, higher literacy does not necessarily reduce phishing susceptibility among African immigrant students or US-born students. While these populations may be more cautious, they still provided disclosures. These findings underscore the need for culturally and linguistically tailored cybersecurity interventions that account for both technical skills and social trust dynamics.

## Acknowledgments

This research was supported by the National Science Foundation under Award No. 2210185. We thank four anonymous reviewers and the revision editor for their constructive feedback, Naomie Ilunga, Samuel Bisimwa for interpreting during the workshops, Enid Ortiz, Emily Cruz, and the IRC for help in recruiting the refugee population. We also thank Mohd Sabra, Jaden Nola, and Anette Wangechi for initial help with implementing the study.

## A Digital Literacy Workshop

- **Day 1: Session 1**
  - Conducted Pre-assessment (Demographic questionnaire, Leap-Q questionnaire, *vINCI* digital literacy questionnaire, Digital Device Usage pre-test)
  - Powerpoint presentation on how to operate smartphones, laptop
  - Actionable items:
    * conducting basic operations such as call a friend
    * type a text
    * type on the keyboard
- **Day 2: Session 2**
  - Recap of Day 1/Session 1
  - Powerpoint presentation on what is PII, phishing, how to detect scams, password creation and protection
  - Actionable items:
    * participants downloaded Microsoft Outlook app
    * logged in to their assigned email address
    * participants sent research team an email
  - Conducted Digital Device Usage post-test

### A.1 Demographic Questionnaire

- **Age:**
  - Under 18
  - 18 - 24
  - 25 - 34
  - 35 - 44
  - 45 - 54
  - 55 - 64
  - 65 - 74
  - 75 - 84
  - 85 or older
- **Ethnicity:**
  - White
  - Black or African American
  - Asian
  - Hispanic or Latino
  - Other
- **Country of Origin**:
- **Languages Spoken**:
- **How well do you speak English?**
  - Cannot speak
  - Speak a little bit
  - Speak very well
- **Do you have family in your home country?**
  - Yes
  - No
- **How often are you in touch with family in your home country?**
  - Definitely yes
  - Probably yes
  - Might or might not
  - Probably not
  - Definitely not
- **At what age did you leave your country?**
  - Under 18
  - 18–24
  - 25–34
  - 35–44
  - 45–54
  - 55–64
  - 65–74
  - 75–84
  - 85 or older
- **Gender:**
  - Male
  - Female
  - Transgender
  - Other
  - Prefer not to respond
- **Length of residence in the US:**
  - Less than 1 year
  - 1–2 years
  - Less than 5 years
  - Less than 10 years
  - 10–20 years
  - 20–30 years
  - More than 30 years
- **Have you lived anywhere other than Wichita?**
  - Yes (Open-ended)
  - No
- **Marital status:**
  - Married
  - Widowed
  - Divorced
  - Separated
  - Never married
- **Are you living with your family in the US?**
  - Yes (Open-ended household size)
  - No
- **Do you have children?**
  - Yes (Open-ended number)
  - No
- **Age of children:** (Open-ended)
- **Are your children in school?**
  - Yes
  - No
  - Not school age
  - Graduated from high school
  - Dropped out of school
- **Languages you speak:** (Open-ended)
- **Languages spoken at home:** (Open-ended)
- **How well do you speak English?**
  - Cannot speak English
  - Speak a little
  - Speak well but have trouble understanding
  - Speak and understand very well

- **How well can you read English?**
  - Extremely well
  - Very well
  - Moderately well
  - Slightly well
  - Not well at all
- **Employment status:**
  - Employed full time
  - Employed part time
  - Unemployed looking for work
  - Unemployed not looking for work
  - Retired
  - Student
  - Disabled
- **Educational qualifications:**
  - Less than high school
  - High school graduate
  - Some college
  - 2-year degree
  - 4-year degree
  - Professional degree
  - Doctorate
- **Do you own the following items?**
  - Phone (landline)
  - Desktop/laptop computer
  - Non-smartphone
  - Smartphone
  - Tablet computer
  - Internet connection
- **Internet provider:**
  - AT&T
  - Verizon
  - Cox
- **Internet speed:**
  - Extremely fast
  - Somewhat fast
  - Average
  - Somewhat slow
  - Extremely slow

## A.2 Language Experience and Proficiency Questionnaire (LEAP-Q)

- **Please list all the languages you know in order of dominance:** _____
- **Please list all the languages you know in order of acquisition (your native language first):** _____
- **Please list what percentage of the time you are currently and on average exposed to each language.**
  - Language 1: ___%
  - Language 2: ___%
  - Language 3: ___%
  - Language 4: ___%
  - Language 5: ___%
- **When choosing to read a text available in all your languages, in what percentage of cases would you choose each language?**
  - Language 1: ___%
  - Language 2: ___%
  - Language 3: ___%
  - Language 4: ___%
  - Language 5: ___%
- **When choosing a language to speak with a person equally fluent in all your languages, what percentage of time would you use each language?**
  - Language 1: ___%
  - Language 2: ___%
  - Language 3: ___%
  - Language 4: ___%
  - Language 5: ___%
- **Please name the cultures with which you identify. Rate your identification with each culture on a scale from 0 (not at all) to 10 (completely):**
  - Culture 1: ___
  - Culture 2: ___
  - Culture 3: ___
- **How many years of formal education do you have?** ___ years
- **What is your highest education level?**
  - Less than high school
  - High school
  - Some college
  - College
  - Professional training
  - Master's degree
  - Doctorate (Ph.D., M.D., J.D.)
- **Date of immigration to the USA (if applicable):** ______
- **Have you ever had any of the following?** (Check all that apply)
  - Vision problem
  - Hearing impairment
  - Language disability
  - Learning disability

### A.2.1 Language-Specific Questions.

- **Language:** _____
- **This is my ___ language.**
- **Age milestones:**
  - Began acquiring: ___
  - Became fluent: ___
  - Began reading: ___
  - Became fluent in reading: ___
- **Time spent in different environments where this language is spoken:**
  - Country: ___ years, ___ months
  - Family: ___ years, ___ months
  - School/Work: ___ years, ___ months
- **Proficiency (0–10):**
  - Speaking: ___
  - Understanding: ___
  - Reading: ___
- **How did you learn this language? Rate each method on a scale from 0 to 10:**
  - Interacting with friends: ___

- Interacting with family: ___
- Watching TV: ___
- Reading: ___
- Language tapes/self-instruction: ___
- Listening to the radio: ___
- **Current exposure to this language (percentage):**
  - Interacting with friends: ___
  - Listening to music/radio: ___
  - Interacting with family: ___
  - Reading: ___
  - Watching TV: ___
- **How much of a foreign accent do you think you have in this language? (0–10):** ___
- **How often are you identified as a non-native speaker based on your accent? (0–10):** ___

### A.3 Digital Skills Questionnaire (vINCI)

- **Devices owned:** (Check all that apply)
  - Desktop Computer
  - Laptop
  - Smartphone
- **Digital Competence:**
  - I can look for information online using a search engine. [Yes/No]
  - I can communicate using email or SMS. [Yes/No]
- **Safety:**
  - I know not to share private information online. [Yes/No]
  - I can protect my devices using antivirus software. [Yes/No]
- **Communication:**
  - I can communicate with others using mobile phones, Voice over IP (e.g., Skype), email, or chat—using basic features (e.g., voice messaging, SMS, send and receive emails, text exchange). [Yes/No]
  - I can share files and content using simple tools. [Yes/No]
  - I know I can use digital technologies to interact with services (e.g., governments, banks, hospitals). [Yes/No]
  - I am aware of social networking sites and online collaboration tools. [Yes/No]
  - I understand that using digital tools requires following certain communication rules (e.g., commenting, sharing personal information). [Yes/No]
- **Content Creation:**
  - I can produce simple digital content (e.g., text, tables, images, audio files) in at least one format using digital tools. [Yes/No]
  - I can make basic edits to content produced by others. [Yes/No]
  - I know that content can be covered by copyright. [Yes/No]
  - I can apply and modify simple functions and settings of software and applications (e.g., changing default settings). [Yes/No]
- **Problem Solving:**
  - I can find support and assistance when a technical problem occurs or when using a new device, program, or application. [Yes/No]
  - I know how to solve some routine problems (e.g., close a program, restart the computer, reinstall/update a program, check the internet connection). [Yes/No]
  - I know that digital tools can help me solve problems, but I am also aware of their limitations. [Yes/No]
  - When confronted with a technological or non-technological problem, I can use the digital tools I know to solve it. [Yes/No]

### A.4 Digital Device Usage Questionnaire

This questionnaire is inspired by the security attitudes questionnaire, however, we created our own version for the refugee population. The questionnaire was created in English and translated into Kiswahili.

- **Instructions:** Lazima useme kati ya 1 na 5 ikiwa unakubali au hukubaliani.
  - 1: Strongly disagree (kutokubaliana kabisa)
  - 2: Somewhat disagree (kutokubaliana kwa kiasi fulani)
  - 3: Neither disagree nor agree (sikubaliani wala kukubaliana)
  - 4: Somewhat agree (kukubaliana kwa kiasi fulani)
  - 5: Strongly agree (kukubaliana sana)
- **I feel comfortable operating a smartphone:**
  - 1, 2, 3, 4, 5
- **I feel comfortable making a phone call:**
  - 1, 2, 3, 4, 5
- **I feel comfortable operating a desktop or a laptop:**
  - 1, 2, 3, 4, 5
- **I know the difference between a phone call that is safe to accept and a phone call that is not safe to accept:**
  - 1, 2, 3, 4, 5
- **I know the difference between information that should and should not be shared during a phone call:**
  - 1, 2, 3, 4, 5

### A.5 Statistical Tests - Workshop Questionnaires

**Table 6: Paired Two Sample t-Test for Means Results of Digital Device Usage Questionnaire.**

| Statistic | Pre-test | Post-test |
|---|---|---|
| Mean | 2.816598 | 3.585198 |
| Variance | 0.437419 | 0.565391 |
| Observations | 5 | 5 |
| Pearson Correlation | | 0.8542072 |
| Hypothesized Mean Difference | | 0 |
| Degrees of Freedom (df) | | 4 |
| t Stat | | -4.390838766 |
| P(T ≤ t) one-tail | | **0.005887456** |
| t Critical one-tail | | 2.131846786 |
| P(T ≤ t) two-tail | | **0.011774913** |
| t Critical two-tail | | 2.776445105 |

**Table 7: Mean Pre-test and Post-test scores for the digital device usage questionnaire (Refer to Section 4.1.4.)**

| Item | Pre | Post |
|---|---|---|
| 1 | 3.0625 | 3.65583 |
| 2 | 3.833 | 4.52083 |
| 3 | 2.14583 | 2.416 |
| 4 | 2.64583 | 3.64583 |
| 5 | 2.39583 | 3.6875 |
| **Overall** | **2.8167** | **3.583** |

# B Phishing Scenarios

## B.1 NIST Phish Scale

We applied the NIST Phish Scale [10] developed by Dawkins and Jacobs to systematically assess the human detection difficulty of phishing email messages. To adapt the NIST Phish Scale for our context, we made several contextual modifications while maintaining the core methodology. The original Phish Scale relies on two dimensions, observable cues and premise alignment, which we retained. However, to reflect the higher education environment, we adjusted the interpretation of premise alignment to account for academic communication patterns. For example, messages referencing common university services such as course registration, student accounts, or faculty communication were considered to have stronger alignment with students' expectations. Similarly, the classification of cues was tailored to include markers typical in academic emails, such as institutional logos, references to campus portals, or mentions of deadlines tied to academic calendars. These adjustments allowed the Phish Scale to be applied consistently, while ensuring that the resulting detection difficulty ratings accurately reflected the specific characteristics of phishing attempts targeting university communities. Overall, premise alignment was defined differently for the three participant groups. For students, alignment reflected references to coursework, registration, student services, administrative processes, campus service announcements. For refugee participants, premise alignment was defined in relation to communications they are most likely to receive in resettlement contexts, such as messages about immigration status, housing services, healthcare access, or employment, which could strongly influence their expectations and perceived legitimacy of a phishing message. All 48 emails (Tables 8 and 12) from the context of groups B and C were rated *Very Difficult*, among which 13 of them were rated as *Moderately Difficult* from the context of Group A.

Table 8: Details for each phishing scenario: Scenario ID, Email Subject, Body, $2 \times 2$ **Manipulation, and Total Compromised Cases** ($N_c$).

| ID | Email Subject | Email Body | $2 \times 2$ | $N_c$ |
|---|---|---|---|---|
| B01 | RESUME: Update Needed | Employers are taking notice of your resume. Reply with your Social Security number for hiring. | PC-MS | 0 |
| B02 | ACT NOW: Job Application | You're almost done with your application. Enter bank account details here. | PC-MS | 0 |
| B03 | OVERTIME BONUS | Your pay has increased from last week. Provide your bank account details now. | PC-MS | 0 |
| B04 | JOB OFFER | Companies are reaching out with job offers. Reply with your Social Security number. | PC-MS | 0 |
| B05 | CLAIM YOUR BENEFITS | We need more details for your benefits. Enter your birthdate and Social Security number. | PC-MS | 0 |
| B06 | HIRING: Information Needed | Best wishes on your new role. Send bank account details to get your paycheck. | PC-MS | 0 |
| B07 | HIRING UPDATE | Wichita State University is processing your hiring. Reply with your Social Security number. | PC-MS | 0 |
| B08 | UPDATE NEEDED: Benefits | Your promotion is being processed. Provide your Social Security number to advance it. | PC-MS | 0 |
| B09 | CONGRATULATIONS: Promotion | Congratulations on your promotion! Respond with your bank account details. | PC-MS | 0 |
| B10 | HIRING: Update | Wichita State University requires your birthdate for hiring confirmation. Reply with your birth date. | PC-MS | 1 |
| B11 | UPDATE INFORMATION | Your date of birth was not included in your job application. Kindly respond with it before the deadline. | PC-MS | 1 |
| B12 | CONGRATULATIONS: Benefits | Well done on earning new healthcare benefits. Enter your birthdate in reply to complete the process. | PC-MS | 0 |
| G01 | Free Rent: CLAIM NOW | WSU Housing is giving away free rent for one month. Reply with your bank account details. | C-MS | 0 |
| G02 | APPLY: Healthcare | WSU Health Services is offering affordable insurance. Respond with your birthdate to be considered. | C-MS | 0 |
| G03 | CLAIM YOUR SCHOLARSHIP | Wichita State University is offering a scholarship. Enter your birthdate to claim it. | C-MS | 1 |
| G04 | APPLY: Free Smartphone | WSU is giving away smartphones. Reply with your Social Security number to claim. | C-MS | 0 |
| G05 | CAR OFFER: Reply First | WSU is offering a Honda automobile. Apply now with your Social Security number. | C-MS | 0 |
| G06 | FINAL NOTICE: Rent | This month's rent payment is canceled. Reply with your bank account details for a refund. | C-MS | 0 |
| G07 | UPDATE: Your Award | The award you won is on its way. Enter your birthdate to arrange pickup. | C-MS | 1 |

Table 8: Scenario Details (continued)

| ID | Email Subject | Body | 2×2 | Nc |
|---|---|---|---|---|
| G08 | PAYMENT DUE | WSU Dining is providing free meals for a month. Reply with your birthdate to claim. | C-MS | 1 |
| G09 | ACT NOW: Housing | Your housing application is incomplete. Submit your date of birth today. | C-MS | 0 |
| G10 | WSU GIVEAWAY | WSU is giving away a $500 gift card. Apply now with your Social Security number. | C-MS | 0 |
| G11 | CHECK VISA STATUS | Your VISA with US Immigration Services was delayed. Reply with your birthdate to expedite it. | C-MS | 0 |
| G12 | FREE TRANSPORTATION | WSU is offering two months of free transportation. Reply with your date of birth to claim. | C-MS | 1 |
| P01 | Debt Outcome | WSU records show you may qualify for debt forgiveness. Click the link to check. | C - DS | 3 |
| P02 | Payment Due | WSU is requesting payment for bills last month. Click the link to view the bill. | C - DS | 5 |
| P03 | New Account Activity | Wichita State University detected unusual login activity. Click the link to modify your password. | C - DS | 2 |
| P04 | Electricity Bill | WSU Housing is disconnecting tenants' electricity. Click to find out if you're impacted. | C - DS | 1 |
| P05 | Credit Card Bill | Your Student credit card balance is due. Click to pay the bill. | C - DS | 3 |
| P06 | Tuition Update | Records from Wichita State University show unpaid tuition. Click to view your debt. | C - DS | 2 |
| P07 | Health Services Balance | Payment required for Wichita State University Health Services. Click to view the balance. | C - DS | 3 |
| P08 | Credit Card Activity | Activity detected on your credit card. Click to confirm. | C - DS | 1 |
| P09 | Unfinished Payment | WSU shows an outstanding payment. Click to view details. | C - DS | 3 |
| P10 | Lost Package | Wichita State University Police are searching for your lost item. Click the claim number to learn more. | C - DS | 0 |
| P11 | Childcare Update | Childcare payment is required at WSU's Child Development Center. To view the balance, click here. | C - DS | 1 |
| P12 | Student Dining Payment | Student Dining records show that you owe $200.14. To view payment options, click this link. | C - DS | 0 |
| Y01 | Insurance Offer | Wichita State University extends its health insurance offer to you. Click the link to send your bank routing number. | PC - DS | 0 |
| Y02 | Verify your Bank Account | WSU is verifying your bank account. Re-enter information by clicking this link. | PC - DS | 2 |
| Y03 | New Job Posting | You have been hired by WSU for a new position. Submit your Social Security number using this link. | PC - DS | 2 |
| Y04 | WSU Job Offer | WSU is processing your resume review payment. Provide payment information by clicking this link. | PC - DS | 0 |
| Y05 | Bank Account Information | Your bank account information needs updating. Click the link to update it. | PC - DS | 1 |
| Y06 | WSU Hiring | WSU is hiring you for a managerial role. Create an account using your Social Security number here. | PC - DS | 0 |
| Y07 | Expired Account Information | WSU records show invalid payment details. Click the link to update your bank account details. | PC - DS | 3 |
| Y08 | Complete Application | You did not enter your Social Security number in your WSU debt relief application. Submit it here. | PC - DS | 0 |
| Y09 | WSU Hiring | WSU Human Resources is offering a payment plan. Click the link to verify your bank details. | PC - DS | 1 |
| Y10 | Benefits at WSU | WSU has reduced your payments. Confirm with your bank account information by clicking here. | PC - DS | 0 |
| Y11 | Wichita State University Job Notice | Wichita State University overpaid you for your role. Use this link to send money to WSU. | PC - DS | 0 |
| Y12 | Update Your Information | WSU records reveal your bank account is inaccurate. Click here to change your routing number. | PC - DS | 1 |

## C Debriefing Email

[Greeting]

You took part in a cybersecurity for all study in 2024-2025. As part of the study, researchers at Wichita State University sent you suspicious messages to learn how people respond to potential online threats. These messages were part of an experiment designed to simulate real scam communications. No personal information was collected during the study, only how you interacted with the messages. The purpose was to study how people respond to social engineering threats.

We would like to invite you to join us for a meeting to better understand your cybersecurity and privacy practices. Meetings can be held online or in-person. Participation is completely voluntary. If you decide to participate, you will be asked a few questions, and your answers will be recoded anonymously. You will then be educated about the results of the study, and you will have a chance to ask the researchers questions about the study. The Meeting Sessions will start [Date] until [Date]. Please sign up with [this link].

You also have the option to exclude yourself from the study data if you reply by end of Date. Participation is voluntary, and your data has already been anonymized. No matter your choice, you won't receive any more messages from us. You can opt-out [here].

If you have any questions, please do not hesitate to ask. Thank you. [Sign-off]

## D Study Phase III

Phase III is an ongoing qualitative semi-structured follow-up with participants in the phishing study. Participants were invited to a voluntary follow-up interview when we sent them debriefing emails and compensates an additional $15. Although the full analysis of the qualitative follow-up is beyond the scope of this paper, we provide a table of analyzed responses from the follow-up interviews we have conducted so far in Table 9. Note that responses provided by PID "A1" are in the form of replies to the phishing emails.

Table 9: Phase III Qualitative Responses by Participant.

| PID | Action | Follow-up Responses | Cybersecurity Understanding |
|---|---|---|---|
| C1 | Read email | "I mean if I'm getting it's a personal e-mail then it's basically through messenger or text message. Other than that my e-mail is only used for online shopping or work. If I don't recognize it, then 9 times out of 10 it's a scam, so I don't bother." | "I wish I could do better because I'm sure at some point it's been compromised, but I do get regular credit checks. Informational things on the Internet (e.g., TikTok) are really helpful for avoiding scams." |
| C2 | PII disclosure | "I had an email account taken over and locked out. They got all the information, so I had to change bank accounts and create a new email." | "Password and face recognition notify me if someone tries to break into my laptop or phone." |
| C3 | Read email | "My algorithm deletes unknown emails into junk, or I report or delete it." | "I unsubscribe from any promotional email." |
| C4 | Click link | "Urgency, car insurance or package delivered. I don't get packages often, so I know those are scams." | "Older people with low digital literacy are more vulnerable. AI and deepfakes will make scams harder to detect." |
| C5 | PII disclosure | "Not connected to my job but insisting needed my information." | "We need more awareness and more resources so people can be protected from scams." |
| C6 | Read email | "Scammed for job offer, so I know to look out for job offers and tracking links." | "All unknown messages go to spam. I use VPNs to protect myself." |
| B1 | Read email | "Been scammed through Snapchat when I was 18, so I know how to recognize scams and look for unfamiliar numbers." | "Just like Title IX training, students should do mandatory cybersecurity training at universities." |
| B2 | Read email | "Asking for my password, pretending to be bank accountant." | "Install software to detect scams." |
| B3 | No action | "Use a burner email but now I get more emails through my primary email." | "Use adblocks." |
| A1 | Replied | "I have received the Two messages from Louise Ndembe and Antoinette Bwanga asking me for My Social Security Number and the other telling me about the problem of My Password. I answered all 2 "NEGATIVE" because all that are my secrets" | "I'm very happy to be one of the participants in the workshop you organized and conducted. I'm really grateful to be able to learn how to open a Outlook account, manage passwords as well as operating a cellphone. I'm pretty sure, the knowledge and skills I learned will help me in my current and future endeavors. Next time, don't forget if you have any other workshop in which you think I can benefit something. God bless you" |

# E Quantitative and Statistical Analysis

Below we provide additional quantitative analyses (Figure 4, Table 10, Table 12,Table 18, Table 19), as well as statistical tables (Table 13, Table 11, Table 17, Table 15, Table 16, Table 14) for Section 5. Statistically significant results ($p < .05$) in this section are highlighted in bold and green.

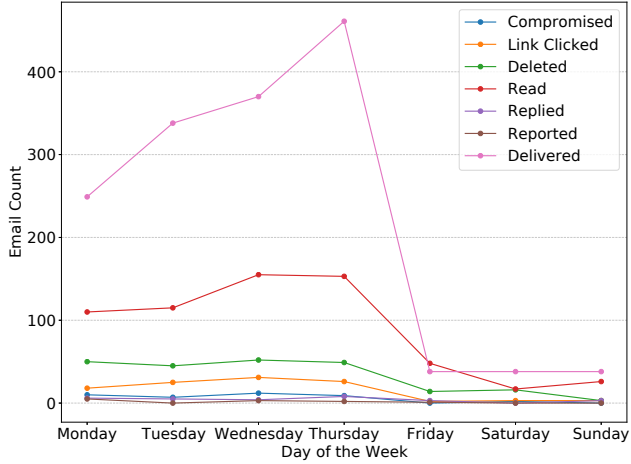

Figure 4: Combined Trends in Phishing Metrics Across the Days of the Week Across Both Studies.

Table 10: Group A Counts by Country and $2 \times 2$ Manipulation.

| Country | $2 \times 2$ | Click | Deleted | Read | Replied | Delivered |
|---|---|---|---|---|---|---|
| Burundi | C-DS | - | - | 1 (33.3%) | - | 3 (100%) |
| Burundi | PC-DS | - | - | 2 (66.7%) | - | 3 (100%) |
| Burundi | C-MS | - | - | 2 (66.7%) | - | 3 (100%) |
| Burundi | PC-MS | - | - | 3 (100%) | - | 3 (100%) |
| DRC | C-DS | - | 3 (6.3%) | 14 (29.2%) | - | 48 (100%) |
| DRC | PC-DS | - | 3 (6.3%) | 13 (27.1%) | - | 48 (100%) |
| DRC | C-MS | - | 4 (8.3%) | 12 (25.0%) | - | 48 (100%) |
| DRC | PC-MS | - | 3 (6.3%) | 14 (29.2%) | - | 48 (100%) |
| Malawi | C-DS | 2 (66.7%) | - | 3 (100%) | 2 (66.7%) | 3 (100%) |
| Malawi | PC-DS | 1 (33.3%) | - | 3 (100%) | 2 (66.7%) | 3 (100%) |
| Malawi | C-MS | 1 (33.3%) | - | 3 (100%) | 3 (100%) | 3 (100%) |
| Malawi | PC-MS | - | - | 3 (100%) | 3 (100%) | 3 (100%) |
| SS | C-DS | - | - | - | - | 3 (100%) |
| SS | PC-DS | - | - | - | - | 3 (100%) |
| SS | C-MS | - | - | - | - | 3 (100%) |
| SS | PC-MS | - | - | - | - | 3 (100%) |

*DRC and SS represents Democratic Republic of the Congo and South Sudan respectively.

Table 11: Summary of Mann-Whitney U Test Results for *Compromised* Responses in Study 2 (Groups B and C).

| Factor | Groups Compared | N (per group) | U-statistic | P-value |
|---|---|---|---|---|
| Legitimacy | DS vs. MS | 652 | 221,680.0 | **< 0.001** |
| Linguistic | C vs. PC | 652 | 217,768.0 | **0.010** |

Table 12: NIST Phish Scale Ratings by Scenario and Group with Cue, Premise Alignment, and Detection Difficulty Ratings. Values are based on the NIST Phish Scale Methodology for Assessing Phishing Detection Difficulty.

| Scenario Group = | Cue (A,B,C) | Premise (A) | Difficulty (A) | Premise (B&C) | Difficulty (B&C) |
|---|---|---|---|---|---|
| B01 | Few | ● | ★ | ● | ★ |
| B02 | Few | ● | ★ | ● | ★ |
| B03 | Few | ● | ★ | ● | ★ |
| B04 | Few | ● | ★ | ● | ★ |
| B05 | Few | ● | ★ | ● | ★ |
| B06 | Few | ● | ★ | ● | ★ |
| B07 | Few | ○ | ☆ | ● | ★ |
| B08 | Few | ● | ★ | ● | ★ |
| B09 | Few | ● | ★ | ● | ★ |
| B10 | Few | ○ | ☆ | ● | ★ |
| B11 | Few | ● | ★ | ● | ★ |
| B12 | Few | ● | ★ | ● | ★ |
| G01 | Few | ○ | ☆ | ● | ★ |
| G02 | Few | ○ | ☆ | ● | ★ |
| G03 | Few | ○ | ☆ | ● | ★ |
| G04 | Few | ○ | ☆ | ● | ★ |
| G05 | Few | ○ | ☆ | ● | ★ |
| G06 | Few | ● | ★ | ● | ★ |
| G07 | Few | ○ | ☆ | ● | ★ |
| G08 | Few | ○ | ☆ | ● | ★ |
| G09 | Few | ○ | ☆ | ● | ★ |
| G10 | Few | ○ | ☆ | ● | ★ |
| G11 | Few | ○ | ☆ | ● | ★ |
| G12 | Few | ○ | ☆ | ● | ★ |
| P01 | Few | ◐ | ★ | ● | ★ |
| P02 | Few | ◐ | ★ | ● | ★ |
| P03 | Few | ◐ | ★ | ● | ★ |
| P04 | Few | ◐ | ★ | ● | ★ |
| P05 | Few | ● | ★ | ● | ★ |
| P06 | Few | ◐ | ★ | ● | ★ |
| P07 | Few | ● | ★ | ● | ★ |
| P08 | Few | ◐ | ★ | ● | ★ |
| P09 | Few | ◐ | ★ | ● | ★ |
| P10 | Few | ● | ★ | ● | ★ |
| P11 | Few | ◐ | ★ | ● | ★ |
| P12 | Few | ◐ | ★ | ● | ★ |
| Y01 | Few | ◐ | ★ | ● | ★ |
| Y02 | Few | ◐ | ★ | ● | ★ |
| Y03 | Few | ◐ | ★ | ● | ★ |
| Y04 | Few | ◐ | ★ | ● | ★ |
| Y05 | Few | ● | ★ | ● | ★ |
| Y06 | Few | ◐ | ★ | ● | ★ |
| Y07 | Few | ◐ | ★ | ● | ★ |
| Y08 | Few | ◐ | ★ | ● | ★ |
| Y09 | Few | ◐ | ★ | ● | ★ |
| Y10 | Few | ◐ | ★ | ● | ★ |
| Y11 | Few | ◐ | ★ | ● | ★ |
| Y12 | Few | ◐ | ★ | ● | ★ |

**Premise Alignment**
● Strong
○ Weak
◐ Medium

**Detection Difficulty**
★ Very difficult
☆ Moderately difficult

Table 13: Logistic Regression Results by Phishing Responses and Manipulation for all Groups (A, B and C) Across Studies.

| Outcome | Predictor | B | SE | z | p | OR | OR [95% CI] |
|---|---|---|---|---|---|---|---|
| Compromised | C-DS | -2.705 | 0.211 | -12.831 | **<0.001** | 0.067 | [0.044, 0.101] |
| | C-MS | -1.846 | 0.545 | -3.387 | **0.001** | 0.158 | [0.054, 0.459] |
| | PC-DS | -0.914 | 0.384 | -2.382 | **0.017** | 0.401 | [0.189, 0.851] |
| | PC-MS | -2.544 | 0.740 | -3.440 | **0.001** | 0.079 | [0.018, 0.335] |
| Click | C-DS | -2.094 | 0.164 | -12.806 | **<0.001** | 0.123 | [0.089, 0.170] |
| | C-MS | -0.656 | 0.270 | -2.429 | **0.015** | 0.519 | [0.305, 0.881] |
| | PC-DS | -0.335 | 0.249 | -1.349 | 0.177 | 0.715 | [0.439, 1.164] |
| | PC-MS | -1.337 | 0.336 | -3.982 | **<0.001** | 0.263 | [0.136, 0.507] |
| Deleted | C-DS | -2.094 | 0.164 | -12.806 | **<0.001** | 0.123 | [0.089, 0.170] |
| | C-MS | 0.507 | 0.213 | 2.381 | **0.017** | 1.660 | [1.094, 2.518] |
| | PC-DS | 0.265 | 0.221 | 1.203 | 0.229 | 1.304 | [0.846, 2.009] |
| | PC-MS | 0.579 | 0.211 | 2.747 | **0.006** | 1.784 | [1.180, 2.697] |
| Read | C-DS | -0.215 | 0.103 | -2.091 | **0.037** | 0.807 | [0.659, 0.987] |
| | C-MS | -0.281 | 0.147 | -1.907 | 0.057 | 0.755 | [0.566, 1.008] |
| | PC-DS | -0.171 | 0.146 | -1.169 | 0.242 | 0.843 | [0.633, 1.123] |
| | PC-MS | -0.193 | 0.146 | -1.316 | 0.188 | 0.825 | [0.619, 1.099] |
| Replied | C-DS | -4.141 | 0.411 | -10.062 | **<0.001** | 0.016 | [0.007, 0.036] |
| | C-MS | 0.414 | 0.532 | 0.777 | 0.437 | 1.512 | [0.533, 4.290] |
| | PC-DS | 0.157 | 0.561 | 0.279 | 0.780 | 1.170 | [0.389, 3.513] |
| | PC-MS | 0.157 | 0.561 | 0.279 | 0.780 | 1.170 | [0.389, 3.513] |
| Reported | C-DS | -5.945 | 1.001 | -5.938 | **<0.001** | 0.003 | [0.000, 0.019] |
| | C-MS | 1.104 | 1.157 | 0.954 | 0.340 | 3.016 | [0.312, 29.122] |
| | PC-DS | 1.104 | 1.157 | 0.954 | 0.340 | 3.016 | [0.312, 29.122] |
| | PC-MS | 1.394 | 1.120 | 1.244 | 0.213 | 4.032 | [0.449, 36.237] |

Table 14: Pairwise Comparisons from Tukey's HSD Post-Hoc Test for *Compromised* Events for Groups B and C.

| Group 1 | Group 2 | Mean Diff. | P-value | 95% CI | Reject |
|---|---|---|---|---|---|
| C − DS + B | C − DS + C | 0.0431 | 0.3149 | [-0.0147, 0.1009] | False |
| C − DS + B | C − MS + B | -0.0282 | 0.8609 | [-0.0896, 0.0332] | False |
| C − DS + B | C − MS + C | -0.0439 | 0.2923 | [-0.1017, 0.0139] | False |
| C − DS + B | PC − DS + B | -0.0282 | 0.8609 | [-0.0896, 0.0332] | False |
| C − DS + B | PC − DS + C | -0.0113 | 0.9990 | [-0.0691, 0.0466] | False |
| C − DS + B | PC − MS + B | -0.0352 | 0.6605 | [-0.0966, 0.0262] | False |
| C − DS + B | PC − MS + C | -0.0493 | 0.1608 | [-0.1071, 0.0085] | False |
| C − DS + C | C − MS + B | -0.0713 | **0.0047** | [-0.1291, -0.0135] | True |
| C − DS + C | C − MS + C | -0.0870 | **< 0.001** | [-0.1409, -0.0330] | True |
| C − DS + C | PC − DS + B | -0.0713 | **0.0047** | [-0.1291, -0.0135] | True |
| C − DS + C | PC − DS + C | -0.0543 | **0.0468** | [-0.1083, -0.0004] | True |
| C − DS + C | PC − MS + B | -0.0783 | **0.0011** | [-0.1361, -0.0205] | True |
| C − DS + C | PC − MS + C | -0.0924 | **< 0.001** | [-0.1463, -0.0384] | True |
| C − MS + B | C − MS + C | -0.0157 | 0.9918 | [-0.0735, 0.0421] | False |
| C − MS + B | PC − DS + B | 0.0000 | 1.0000 | [-0.0614, 0.0614] | False |
| C − MS + B | PC − DS + C | 0.0169 | 0.9871 | [-0.0409, 0.0747] | False |
| C − MS + B | PC − MS + B | -0.0070 | 1.0000 | [-0.0685, 0.0544] | False |
| C − MS + B | PC − MS + C | -0.0211 | 0.9548 | [-0.0789, 0.0367] | False |
| C − MS + C | PC − DS + B | 0.0157 | 0.9918 | [-0.0421, 0.0735] | False |
| C − MS + C | PC − DS + C | 0.0326 | 0.5962 | [-0.0213, 0.0866] | False |
| C − MS + C | PC − MS + B | 0.0086 | 0.9998 | [-0.0492, 0.0665] | False |
| C − MS + C | PC − MS + C | -0.0054 | 1.0000 | [-0.0594, 0.0485] | False |
| PC − DS + B | PC − DS + C | 0.0169 | 0.9871 | [-0.0409, 0.0747] | False |
| PC − DS + B | PC − MS + B | -0.0070 | 1.0000 | [-0.0685, 0.0544] | False |
| PC − DS + B | PC − MS + C | -0.0211 | 0.9548 | [-0.0789, 0.0367] | False |
| PC − DS + C | PC − MS + B | -0.0240 | 0.9138 | [-0.0818, 0.0338] | False |
| PC − DS + C | PC − MS + C | -0.0380 | 0.3892 | [-0.0920, 0.0159] | False |
| PC − MS + B | PC − MS + C | -0.0141 | 0.9958 | [-0.0719, 0.0437] | False |

Table 15: Logistic Regression Results for Phishing Responses by 2 × 2 Manipulations for Groups B and C.

| Outcome | Predictor | B | SE | z | p | OR | OR [95% CI] |
|---|---|---|---|---|---|---|---|
| Compromised | C-DS | -2.532 | 0.212 | -11.94 | **<0.001** | 0.08 | [0.05, 0.12] |
| | C-MS | -1.856 | 0.546 | -3.40 | **0.001** | 0.16 | [0.05, 0.46] |
| | PC-DS | -0.921 | 0.385 | -2.39 | **0.017** | 0.40 | [0.19, 0.85] |
| | PC-MS | -2.555 | 0.740 | -3.45 | **0.001** | 0.08 | [0.02, 0.33] |
| Click | C-DS | -1.967 | 0.169 | -11.65 | **<0.001** | 0.14 | [0.10, 0.19] |
| | C-MS | -0.659 | 0.278 | -2.37 | **0.018** | 0.52 | [0.30, 0.89] |
| | PC-DS | -0.322 | 0.255 | -1.26 | 0.207 | 0.72 | [0.44, 1.20] |
| | PC-MS | -1.297 | 0.339 | -3.83 | **<0.001** | 0.27 | [0.14, 0.53] |
| Deleted | C-DS | -1.996 | 0.171 | -11.70 | **<0.001** | 0.14 | [0.10, 0.19] |
| | C-MS | 0.527 | 0.222 | 2.37 | **0.018** | 1.69 | [1.10, 2.62] |
| | PC-DS | 0.288 | 0.230 | 1.25 | 0.211 | 1.33 | [0.85, 2.09] |
| | PC-MS | 0.625 | 0.219 | 2.85 | **0.004** | 1.87 | [1.22, 2.87] |
| Read | C-DS | -0.123 | 0.111 | -1.11 | 0.268 | 0.88 | [0.71, 1.10] |
| | C-MS | -0.313 | 0.159 | -1.98 | **0.048** | 0.73 | [0.54, 1.00] |
| | PC-DS | -0.199 | 0.158 | -1.26 | 0.208 | 0.82 | [0.60, 1.12] |
| | PC-MS | -0.250 | 0.158 | -1.58 | 0.115 | 0.78 | [0.57, 1.06] |
| Replied | C-DS | -4.388 | 0.503 | -8.72 | **<0.001** | 0.01 | [0.00, 0.03] |
| | C-MS | 0.412 | 0.650 | 0.63 | 0.527 | 1.51 | [0.42, 5.40] |
| | PC-DS | 0.226 | 0.675 | 0.34 | 0.738 | 1.25 | [0.33, 4.71] |
| | PC-MS | 0.000 | 0.711 | 0.00 | 1.000 | 1.00 | [0.25, 4.03] |
| Reported | C-DS | -5.784 | 1.002 | -5.78 | **<0.001** | 0.00 | [0.00, 0.02] |
| | C-MS | 1.105 | 1.157 | 0.96 | 0.340 | 3.02 | [0.31, 29.17] |
| | PC-DS | 1.105 | 1.157 | 0.96 | 0.340 | 3.02 | [0.31, 29.17] |
| | PC-MS | 1.396 | 1.121 | 1.25 | 0.213 | 4.04 | [0.45, 36.32] |

Table 16: Logistic Regression Results for Phishing Responses by English Reading Proficiency (ERP) for Groups B and C.

| Outcome | Predictor | B | SE | z | p | OR | OR [95% CI] |
|---|---|---|---|---|---|---|---|
| Compromised | ERP=4 | -3.607 | 0.262 | -13.79 | **<0.001** | 0.03 | [0.02, 0.05] |
| | ERP=5 | 0.260 | 0.331 | 0.78 | 0.434 | 1.30 | [0.68, 2.48] |
| Click | ERP=4 | -2.693 | 0.172 | -15.64 | **<0.001** | 0.07 | [0.05, 0.09] |
| | ERP=5 | 0.408 | 0.214 | 1.91 | 0.057 | 1.50 | [0.99, 2.29] |
| Deleted | ERP=4 | -2.406 | 0.152 | -15.80 | **<0.001** | 0.09 | [0.07, 0.12] |
| | ERP=5 | 1.195 | 0.176 | 6.80 | **<0.001** | 3.30 | [2.34, 4.66] |
| Read | ERP=4 | -1.071 | 0.096 | -11.13 | **<0.001** | 0.34 | [0.28, 0.41] |
| | ERP=5 | 1.278 | 0.121 | 10.52 | **<0.001** | 3.59 | [2.83, 4.55] |
| Replied | ERP=4 | -5.238 | 0.579 | -9.05 | **<0.001** | 0.01 | [0.00, 0.02] |
| | ERP=5 | 1.432 | 0.632 | 2.27 | **0.023** | 4.19 | [1.21, 14.43] |
| Reported | ERP=4 | -6.340 | 1.001 | -6.34 | **<0.001** | 0.00 | [0.00, 0.01] |
| | ERP=5 | 2.055 | 1.050 | 1.96 | 0.050 | 7.81 | [1.00, 61.19] |

*Uses data combined from Study 2. ERP values 4 and 5 denote that participants can read English very well and extremely well, respectively.

Table 17: Two-Way ANOVA Results for *Compromised* Responses in Study 2 (Groups B and C).

| Source | Sum of Squares | df | F-statistic | P-value |
|---|---|---|---|---|
| 2x2 Manipulation | 0.908 | 3.0 | 10.4161 | **< 0.001** |
| Participant Group | 0.018 | 1.0 | 0.6304 | 0.427 |
| Manipulation × Group | 0.189 | 3.0 | 2.1694 | 0.090 |
| Residual | 37.658 | 1296.0 | | |

**Table 18: Participant Email Interactions by Group and Week for Study #2 (Per-week Count and Percentage, Total Emails $B$ = 142, $C$ = 184). Unread Emails were Excluded from this Table.**

|  | Group B | | | | Group C | | | |
| --- | --- | --- | --- | --- | --- | --- | --- | --- |
| Week: | 1 | 2 | 3 | 4 | 1 | 2 | 3 | 4 |
| Read | 37 (26%) | 37 (26%) | 38 (27%) | 33 (23%) | 112 (61%) | 103 (56%) | 93 (51%) | 98 (53%) |
| Deleted | 11 (8%) | 12 (8%) | 12 (8%) | 12 (8%) | 38 (21%) | 44 (24%) | 40 (22%) | 47 (26%) |
| Replied | 2 (1%) | 0 (0%) | 1 (1%) | 0 (0%) | 4 (2%) | 5 (3%) | 2 (1%) | 5 (3%) |
| Forwarded | 0 (0%) | 0 (0%) | 0 (0%) | 0 (0%) | 2 (1%) | 1 (1%) | 1 (1%) | 0 (0%) |
| Reported | 0 (0%) | 0 (0%) | 1 (1%) | 0 (0%) | 3 (2%) | 2 (1%) | 0 (0%) | 5 (3%) |
| Link Clicked | 7 (5%) | 12 (8%) | 9 (6%) | 8 (6%) | 23 (13%) | 16 (9%) | 15 (8%) | 14 (8%) |
| Compromised | 3 (2%) | 5 (4%) | 2 (1%) | 5 (4%) | 11 (6%) | 6 (3%) | 3 (2%) | 5 (3%) |

**Table 19: Country-wise Event Counts and Percentages by Manipulation for Study 2 with Groups B and C (Percentages Based on Delivered Total per Country).**

| Country | Manipulation | Compromised | Click | Deleted | Read | Replied | Reported | Delivered |
| --- | --- | --- | --- | --- | --- | --- | --- | --- |
| Burundi | C-DS | 1 (16.67%) | 1 (16.67%) | – | 2 (33.33%) | – | – | 2 |
| Burundi | C-MS | – | 1 (16.67%) | – | 1 (16.67%) | – | – | 1 |
| Burundi | PC-DS | – | 1 (16.67%) | 1 (16.67%) | 1 (16.67%) | – | – | 2 |
| Burundi | PC-MS | – | – | 1 (16.67%) | 1 (16.67%) | – | – | 1 |
| DRC | C-DS | 1 (1.25%) | 1 (1.25%) | 5 (6.25%) | 13 (16.25%) | – | – | 21 |
| DRC | C-MS | 1 (1.25%) | 2 (2.50%) | 3 (3.75%) | 8 (10.00%) | – | – | 18 |
| DRC | PC-DS | – | 3 (3.75%) | 5 (6.25%) | 11 (13.75%) | – | – | 20 |
| DRC | PC-MS | – | – | 4 (5.00%) | 9 (11.25%) | – | – | 21 |
| Kenya | C-DS | 2 (0.52%) | 5 (1.29%) | 1 (0.26%) | 8 (2.06%) | – | – | 97 |
| Kenya | C-MS | – | 2 (0.52%) | 3 (0.78%) | 7 (1.81%) | – | 1 (0.26%) | 86 |
| Kenya | PC-DS | 1 (0.26%) | 3 (0.78%) | 3 (0.78%) | 11 (2.86%) | – | – | 94 |
| Kenya | PC-MS | – | – | 5 (1.29%) | 9 (2.32%) | – | – | 97 |
| Rwanda | C-DS | 1 (12.50%) | 1 (12.50%) | 1 (12.50%) | 2 (25.00%) | – | – | 2 |
| Rwanda | C-MS | – | 1 (12.50%) | 1 (12.50%) | 2 (25.00%) | 1 (12.50%) | – | 2 |
| Rwanda | PC-DS | – | – | – | 1 (12.50%) | – | – | 2 |
| Rwanda | PC-MS | – | – | 1 (12.50%) | 1 (12.50%) | – | – | 2 |
| Tanzania | C-DS | 2 (1.56%) | 2 (1.56%) | 2 (1.56%) | 11 (8.59%) | – | – | 16 |
| Tanzania | C-MS | 2 (1.56%) | 4 (3.13%) | 4 (3.13%) | 11 (8.59%) | – | – | 16 |
| Tanzania | PC-DS | 1 (0.78%) | 2 (1.56%) | 2 (1.56%) | 12 (9.38%) | – | – | 16 |
| Tanzania | PC-MS | 1 (0.78%) | 4 (3.13%) | 4 (3.13%) | 12 (9.38%) | 1 (0.78%) | – | 16 |
| Uganda | C-DS | – | – | – | 2 (16.67%) | – | – | 3 |
| Uganda | C-MS | – | 1 (8.33%) | – | 1 (8.33%) | – | – | 3 |
| Uganda | PC-DS | 1 (8.33%) | 1 (8.33%) | – | 2 (16.67%) | – | – | 3 |
| Uganda | PC-MS | 1 (8.33%) | 1 (8.33%) | – | 2 (16.67%) | – | – | 3 |
| US | C-DS | 17 (2.31%) | 30 (4.08%) | 30 (4.08%) | 115 (15.65%) | 4 (0.54%) | 1 (0.14%) | 184 |
| US | C-MS | 1 (0.14%) | 11 (1.50%) | 49 (6.68%) | 94 (12.75%) | 4 (0.54%) | 2 (0.27%) | 184 |
| US | PC-DS | 7 (0.95%) | 20 (2.72%) | 39 (5.32%) | 99 (13.43%) | 5 (0.68%) | 3 (0.41%) | 184 |
| US | PC-MS | – | 7 (0.95%) | 51 (6.95%) | 98 (13.29%) | 3 (0.41%) | 4 (0.54%) | 184 |

Note: The table excludes data from one participant from South Sudan who received an email in every manipulation, but did not interact with the emails.